\documentclass[12pt,prd,tightenlines,preprintnumbers,showpacs,nofootinbib,floatfix,superscriptaddress]{revtex4}
\usepackage{epsfig}
\usepackage{amssymb}

\newcommand{\cC}{{\cal C}}
\newcommand{\cZ}{{\cal Z}}
\newcommand{\Z}{{Z \!\!\! Z}}
\newcommand{\LL}{{I \!\! L}}
\newcommand{\beqn}{\begin{eqnarray}}
\newcommand{\eeqn}{\end{eqnarray}}
\newcommand{\eq}[1]{(\ref{#1})}

\newcommand{\dd}{\mathrm{d}}
\newcommand{\dD}{{\cal D}}
\newcommand{\dual}{\, \mbox{}^{\ast}}

\begin{document}

\preprint{ITEP-LAT/2005-02}
\preprint{HU-EP-05/05}
\preprint{LU-ITP 2005/007}

\title{The compact $Q=2$ Abelian Higgs model in the London limit: \\
vortex--monopole chains and the photon propagator}

\author{M.~N.~Chernodub}\affiliation{ITEP, Bolshaya Cheremushkinskaja 25, Moscow, 117218, Russia}
\author{R.~Feldmann}\affiliation{Institut f\"ur Theoretische Physik and NTZ, Universit\"at
Leipzig, D-04109 Leipzig, Germany}
\author{E.-M.~Ilgenfritz}\affiliation{Institut f\"ur Physik, Humboldt-Universit\"at zu Berlin,
Newtonstrasse 15, D-12489 Berlin, Germany}
\author{A.~Schiller}\affiliation{Institut f\"ur Theoretische Physik and NTZ, Universit\"at
Leipzig, D-04109 Leipzig, Germany}

\begin{abstract}
The confining and topological properties of the compact Abelian Higgs model with
doubly--charged Higgs field in three space-time dimensions are studied.
We consider the London limit of the model.
We show that the monopoles are forming chain-like structures
(kept together by ANO vortices) the presence of which is essential for getting
simultaneously permanent confinement of singly--charged particles and breaking of
the string spanned between doubly--charged particles. In the confinement phase the
chains are forming percolating clusters while in the
deconfinement (Higgs) phase the chains are of finite size. The described picture is
in close analogy with the synthesis of the Abelian monopole and the center vortex
pictures in confining non--Abelian gauge models. The screening properties of the
vacuum are studied by means of the photon propagator in
the
Landau gauge.
\end{abstract}


\date{February 17, 2005}

\maketitle

\section{Introduction}
\label{sec:Introduction}

The Abelian Higgs model with compact gauge fields (cAHM) has
potential applications in various areas of physics. Examples range
from high energy physics~\cite{ref:HEP,Bhanot:1981ug} over condensed
matter physics~\cite{ref:CondMatt,ref:Kleinert} to
neural networks~\cite{ref:Neural} {\it etc.} Recently a
closer relation of this model to the infrared sector of the theory
of strong interaction (QCD) was discussed~\cite{ref:U1Q2PLB}.
The model is also interesting by itself due to its non--perturbative
features which can be treated analytically in some cases.

The non--perturbative features of the cAHM arise due to presence of
topological defects among the excitations of the model. There are two types
of defects: monopoles and Abrikosov-Nielsen-Olesen (ANO) vortices~\cite{ANO}.
The monopoles appear due to the
compactness of the gauge vector field while the existence of vortices
is linked to the compactness of the phase of the Higgs field.
Vortices carry magnetic flux which begins/ends on monopoles/antimonopoles.

The study of the cAHM with doubly--charged Higgs field ($Q=2$ cAHM)
presented below is motivated by the relation of this model
to gluodynamics. In the confinement phase of the $Q=2$ cAHM the
singly--charged external particles must be confined by a linear potential
while the potential between external doubly--charged particles should
flatten at large separation between the charges due to the so-called
string-breaking phenomenon.
This feature makes $Q=2$ cAHM similar to gluodynamics in which the
fundamental charged (static quarks) are confined while the potential between
the adjoint charges (static ``gluons'') is asymptotically screened.

The different effect on singly-- and doubly--charged external particles can
be reconciled~\cite{ref:U1Q2PLB} with monopole dynamics only
if one assumes that monopoles are bound into chains which occupy
the whole volume of the lattice. Indeed, the monopole gas is
known~\cite{ref:Polyakov} to be confining at all distances, and no string
breaking and, consequently, the flattening of the potential is possible.
As a relevant physical example one may cite the compact $U(1)$ gauge model
without matter fields ({\it i.e.} in compact electrodynamics, cQED${}_3$) at
zero temperature where the monopoles form a Coulomb plasma~\cite{ref:Polyakov}.

The other extreme, the vacuum filled {\it only} by a gas of
monopole--anti-monopole bound pairs,
cannot be confining at all~\cite{FrSp80}. This situation is realized in
cQED${}_3$ at high temperatures~\cite{ref:cQED:Binding} as well as in cAHM with
singly--charged dynamical matter fields~\cite{ref:Kleinert,Chernodub:2002ym}.

Therefore, in the case of $Q=2$ cAHM, the monopoles
must form structures different both from the monopole gas and the
magnetically neutral dipole gas. In fact, it was found
numerically~\cite{ref:U1Q2PLB} that in
the presence of a doubly--charged Higgs field the monopoles
actually form chain--like structures. This offers an explanation of both
confinement of singly--charged electric particles and string
breaking for doubly--charged test particles.

The described pattern of charge confinement in $Q=2$ cAHM has a
close analogy with gluodynamics where tight
correlations between Abelian monopoles and center vortices (each in
the respective Abelian projection~\footnote{For reviews of the monopole
and center vortex confining mechanism as well as for a discussion of the
corresponding gauges/projections an interested reader may consult
Ref.~\cite{Review:Monopole} and Ref.~\cite{Review:Vortex}, respectively.})
have been found~\cite{Greensite_3}. It also suggests a natural
mechanism for the formation of monopole sheets (instead of chains in 3D)
in 4D gluodynamics.
For example, in the pure SU(2) gauge model (chosen here for
simplicity of discussion) the Abelian monopoles are defined with the
help of an Abelian gauge, in which the off-diagonal gluons
(originally ignored in the Abelian projection) play the role of the
{\it doubly--charged} matter fields coupled minimally to the leading
diagonal gluons. The presence of the doubly--charged dynamical matter
leads to the formation of the vortices carrying magnetic
flux~\cite{ref:GubarevAHM}. These ``magnetic'' vortices are similar to
center vortices. The off-diagonal matter fields also may cause, as pointed out
in Ref.~\cite{ref:U1Q2PLB}, the monopole trajectories to be confined inside
sheets.

In this paper we complete the study of the cAHM in three space-time dimensions
along the direction outlined in Ref.~\cite{ref:U1Q2PLB}, restricting ourselves
to the London limit. We hope to come back to the model at finite
quartic Higgs coupling on another occasion. We study the structure
of the monopole chains, vortex and monopole cluster properties at various
couplings of the $Q=2$ cAHM model. The (Debye) screening properties of the vacuum
are investigated with the help of the photon propagator.

The structure of this paper is as follows. In
Section~\ref{sec:Model} we describe general properties of the
model on the lattice.
In this Section we also discuss the monopole and vortex structures
and derive their interactions by rewriting the partition function
using the Villain form of the model in the London limit.
The topological nature of the confining properties of the model is pointed out.
In Section~\ref{sec:phase_structure_percolation} we
numerically investigate the phase structure and the percolation properties
of the monopole chains using the model with the cosine form of the action.
Both finite and zero temperatures cases are studied.
In Section~\ref{sec:phase_structure_photon} we
numerically investigate the photon propagator in Landau gauge crossing
the phase transition from the confined to the Higgs phase.
Our conclusions are discussed in Section~\ref{sec:Conclusions}.

\section{The model}
\label{sec:Model}

\subsection{Lattice formulation}
\label{subsec:Lattice_formulation}

We consider the three-dimensional Abelian Higgs model
with compact gauge fields $\theta_{x,\mu} \in [-\pi,\pi)$ which are defined
on links $\{x,\mu\}$. Let the scalar Higgs field, $\Phi_x \in \mathbb{C}$,
be doubly--charged and written in the form
\beqn
\Phi_x= \rho_x {\rm e}^{i \varphi_x}
\eeqn
where $\rho_x$ is its modulus (radial part) and $\varphi_x$ its phase.
The action of the model is
\beqn
  S = - \beta \sum_P \cos\theta_P
            - \kappa \sum_{x,\mu} \rho_x \rho_{x+\hat{\mu}}
        \cos(- \varphi_x
      - 2\, \theta_{x,\mu} + \varphi_{x+\hat{\mu}})
        +\sum_x (\rho_x^2 + \lambda (\rho_x^2-1)^2) \,,
  \label{eq:action}
\eeqn
where $\theta_P$ is the plaquette angle. The parameter $\beta$ is proportional
to the inverse gauge coupling squared, $\beta=1/(a\, g^2)$, $\kappa$ is the
so-called hopping parameter and $\lambda$ the quartic Higgs self coupling.

In a special limit of the model -- the so-called ``London limit'' --
the coupling $\lambda$ is infinite and therefore the radial mode of the Higgs field
is frozen to unity. In this paper we study this
limit in detail and relegate the more general case of the radially active model to
another paper. In the London limit Eq.~\eq{eq:action} reduces to
\beqn
  S_{\rm London}[\theta] =
    - \beta \sum_P \cos\theta_P
    -\kappa \sum_{x,\mu} \cos(- \varphi_x
    - 2\, \theta_{x,\mu} + \varphi_{x+\hat{\mu}})\,.
  \label{eq:London-action}
\eeqn

The phase structure of the reduced model
can be sketched starting from the following
considerations~\cite{ref:HEP,Bhanot:1981ug}.
At vanishing hopping parameter, $\kappa=0$, the model~(\ref{eq:London-action})
reduces to the pure compact Abelian gauge theory.
On the symmetric lattices this model
is confining at any coupling $\beta$ due to the presence of the monopole
plasma~\cite{ref:Polyakov}. Thus the low-$\kappa$ region of the phase diagram
corresponds to the confining phase (``confining region'').
At large values of $\kappa$ (called the ``Higgs region'') the model reduces to
the $\Z_2$ gauge model. Indeed, at very large $\kappa$ we get the constraint
$\varphi_{x+\hat{\mu}} - 2 \, \theta_{x,\mu} - \varphi_x = 2\pi \, l_{x,\mu}$
with $l_{x,\mu}\in \Z$. However, in the unitary gauge, all
$\varphi_x =0$, and the above constraint along with the compactness condition
for the gauge field, $\theta_{x,\mu} \in [-\pi,\pi)$, gives two possible
solutions: $\theta_{x,\mu}= 0$ and $\theta_{x,\mu}= \pi$. One can rewrite the
model of Eq.~\eq{eq:London-action}
in this gauge as the $\Z_2$ gauge model for the variables
$z_{x,\mu} = e^{i \theta_{x,\mu}} = \pm 1$. The $\Z_2$ gauge model in three
dimensions has a second order phase transition at the critical coupling
$\beta^{\Z_2}_c \approx 0.7613$~\cite{Z2phase}.
A transition line separates the confinement
phase [closer to origin, $(\beta,\kappa)=(0,0)$] from the Higgs phase
[closer to the point  $(\beta,\kappa)=(\infty,\infty)$].

\begin{figure*}[!htb]
\begin{center}
\includegraphics[scale=0.8,clip=true]{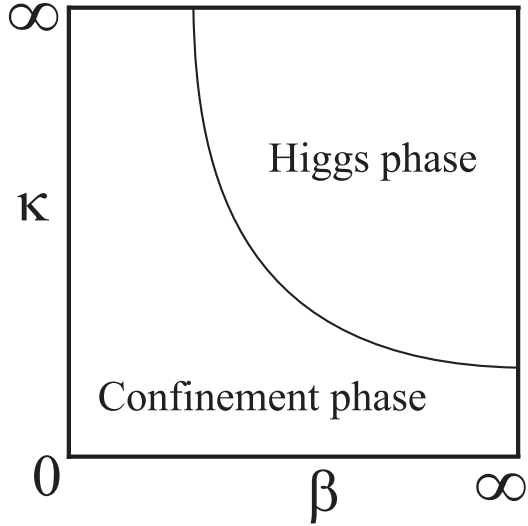}
\end{center}
\vskip -5mm
\caption{The phase diagram of the 3D Q=2 Abelian Higgs model in the London limit.}
\label{fig:phase_diagram_general}
\end{figure*}

At zero value of the coupling constant $\beta$ the model is trivial.
At large $\beta$ the model reduces to the three dimensional $XY$-model due
to the constraint for the plaquette variable,
$\theta_P = 2 \pi n_P$, with $n_P \in \Z$.
This constraint is solved in the form
$\theta_{x,\mu} = - \chi_{x+\hat\mu} + \chi_{x} + 2 \pi l_{x,\mu}$,
where $\chi_{x} \in [-\pi,\pi)$ is a compact scalar field.
Then a re-scaling of the spin field $\chi_x$ by a factor of 2 gives us
the $XY$-model which is known to possess a second order phase
transition~\cite{XYphase} at $\kappa^{XY}_c =0.45420(2)$
between a symmetric and a Higgs phase. The schematic view of the phase
diagram is shown in Figure~\ref{fig:phase_diagram_general}.

{}From our previous studies in the London limit of the $Q=1$ cAHM
at $\beta=2.0$~\cite{Chernodub:2002ym} we know that there
is no transition but the crossover from the
``confining region'' to the ``Higgs region'' is signalled by a
rapid drop of the monopole density. In addition, a string breaking phenomenon
affecting fundamental test charges (being in the same representation as the
fundamentally charged matter fields) has been observed in the Higgs region
whereas they are bound by a linearly rising (string) potential in the
confining region. While the drop of the monopole density signals the onset of
deconfinement,
the anomalous dimension of the photon propagator is turning to zero at the same
crossover $\kappa_c(\beta)$. This accompanies
the transition from the symmetric/confining to the Higgs region, whereas the
observed non-vanishing photon mass at larger $\kappa$ is simply generated by
the Higgs effect.

\subsection{From $Q=1$ to $Q=2$ and beyond}
\label{subsec:From_Q1_to_Q2}

In the compact U(1) model without dynamical matter fields
external static charges $q$ (measured in units of some electric charge $e$)
are permanently confined in three space-time dimensions at zero
temperature~\cite{ref:Polyakov}.
The mechanism is simple: the compact model possesses monopoles which form
a Debye plasma state. If the external charges are immersed into this plasma
moving around a Wilson loop,
monopoles form a sandwich-like structure of magnetic excess charge
of opposite sign on both sides of a certain surface.
This surface is the minimal area surface spanned by the trajectory of the
``electric'' charge. The thickness of this structure is of the order of the
Debye screening length. This leads to a finite excess free energy per unit
area of the double sheet. Thus, oppositely charged external sources at
distance $R$ are confined by a linear potential $V_q(R)$ due to the monopole
plasma.

The situation becomes more complicated when dynamical matter fields with
electric charge $Q~e$ are present where $Q$ is an integer.
In this model -- known as the compact Abelian Higgs model (cAHM) -- monopoles
are not the only topological objects present in the vacuum.
The model additionally possesses so called vortices which carry a certain
magnetic flux. In the general case the total flux carried by the vortex is
$\Phi = 2 \pi/(Q\,e)$. According to the Dirac quantization condition
the monopole carries the charge $m = 2 \pi/e$.
The magnetic flux emitted by an (anti-)monopole is squeezed into the
vortices (the Meissner effect).
The number of vortices attached to a single (anti)monopole is
$Q = m/\Phi$.

If the matter field carries unit charge, $Q=1$, only one vortex is attached
to each (anti)monopole. When the vortices possess a non-zero ``vortex tension''
({\it i.e.} action per length) the monopoles must form monopole--antimonopole
pairs (``monopoliums''
or, in other words, magnetically neutral  dipoles)
confined by a linear potential due to the vortex.
The monopolium has a characteristic size, $R_m$, which is inversely proportional
to this vortex tension. This monopole binding changes now the primordial confining
potential $V_{q,Q}(R)$ between static external electric charges $q~e$.

Firstly, we ignore the Meissner effect and suppose that the vacuum consists of
pairs of size $R_m$. When the monopolium size is small compared to the distance
between the external electric charges, $R_m \ll R$, the influence of any monopole
on the potential $V_{q,Q}(R)$ is almost compensated by that of an antimonopole
belonging to the same pair. Therefore the influence of any
(anti-)monopole on the long-distance part of $V_{q,Q}(R)$ is negligibly
small and confinement is lost. In other words, a vacuum
formed out of dipoles is not confining,
so for $R\gg R_m$ the potential $V_{q,Q}(R) $ becomes flat.
If $R_m \gg R$, the field of a single monopole is practically
unscreened by the presence of other monopoles.
As a consequence, we may expect a piecewise
linear potential due to the usual plasma-like mechanism.
This picture is exactly the string breaking mechanism. While the string breaking
has to appear for external charges of arbitrary strength $q$,
the typical string breaking size might depend on $q$. So we conclude, that
the binding of the monopoles into magnetically neutral pairs may successfully
describe the string breaking in terms of {\it topological} degrees of freedom.

The so far neglected Meissner effect
provides an additional distance scale, which is equal
to the inverse mass of the compact gauge field, $R_\theta = 1/M_\theta$.
This scale defines the width of the physical vortex and also
the characteristic distance at which the magnetic field of the monopole deviates
from the Coulomb behavior due to its squeezing into the vortices. Thus the
presence of the Meissner effect leads to an obvious modification of the condition
that $R$ is larger than the string breaking distance. Instead of $R \gg R_m$
(ignoring the Meissner effect) it is sufficient to have
$R \gg \min(R_m,R_\theta)$ in order to find string breaking.

If the charge of the dynamical matter field is bigger than unity, the situation
becomes more complicated. Consider, for example, the $Q=2$ case.
It is known~\cite{ref:HEP} that
odd-charged external test particles
must be confined in this case while
even-charged ones must show string breaking at suitably large distances.
What structures of monopoles and vortices might correspond to this picture?
If the monopoles would form exclusively neutral dipoles the string breaking
should affect both even- and odd-charged  external sources.
Thus, the monopoles have to form other structures to explain the mentioned behavior.

\begin{figure*}[!htb]
\begin{center}
\begin{tabular}{cc}
\includegraphics[scale=0.7,clip=true]{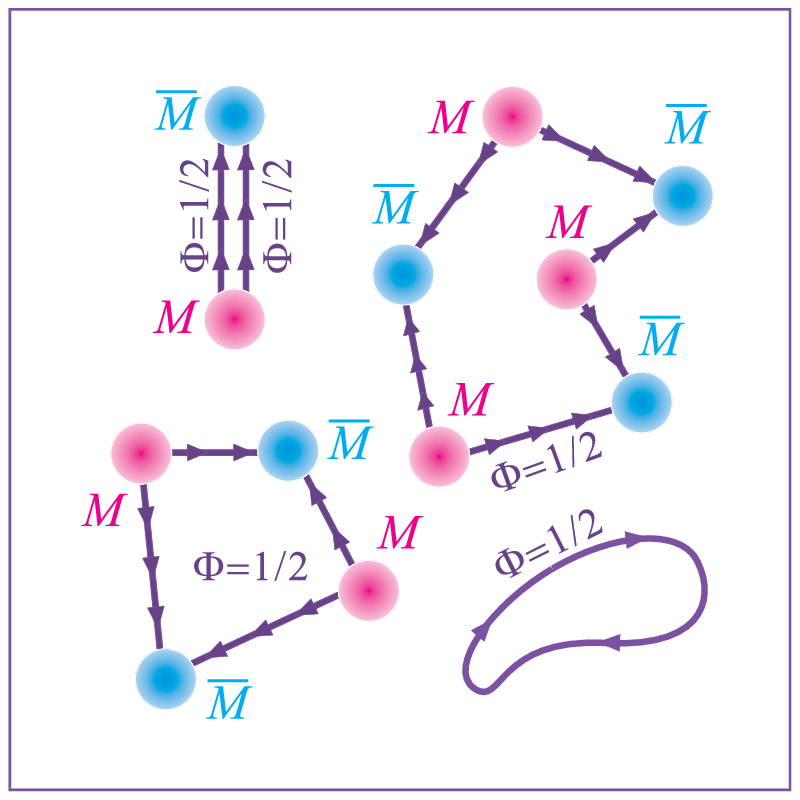} \hspace{1cm} &
\includegraphics[scale=0.7,clip=true]{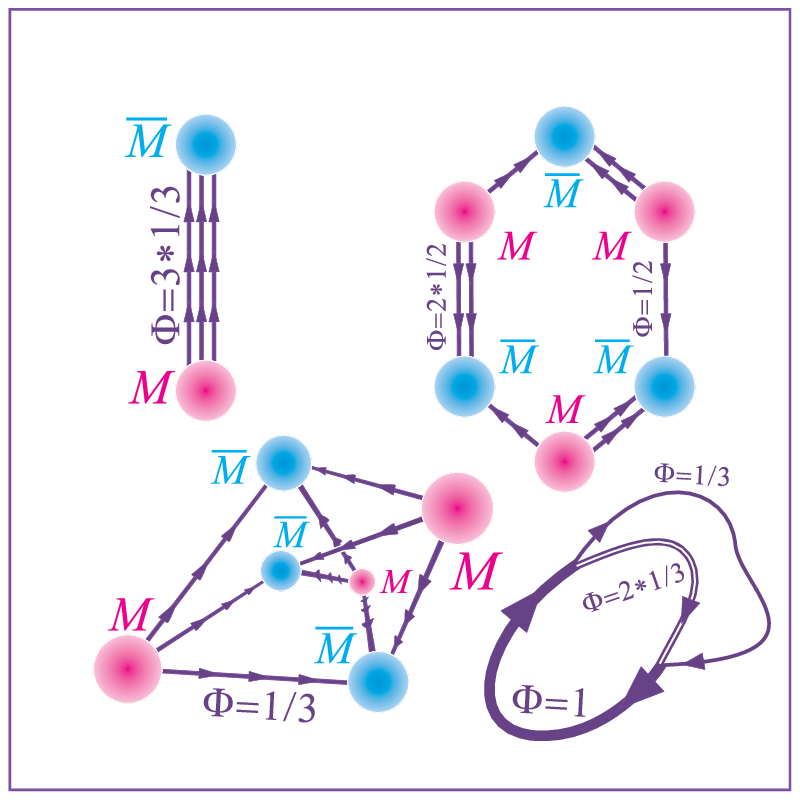} \\
(a) \hspace{1cm} & (b) \\
\end{tabular}
\end{center}
\vskip -5mm
\caption{A schematic view of simplest vortex--monopole and pure vortex configurations
in the cAHM${}_3$ with (a) $Q=2$ and (b) $Q=3$ dynamical Higgs field.}
\label{fig:monopoles_strings_Q2}
\end{figure*}

Since there are $Q$ vortices attached to each (anti-)monopole,
the vortices eventually force the monopoles to form magnetically neutral
dipole-like states as in the $Q=1$ case.
If $Q=2$, in addition monopoles and vortices
may form
also wire-like structures as depicted
in Figure~\ref{fig:monopoles_strings_Q2}(a).
In the case of $Q=3$ the monopoles and vortices may also form
various structures like ``benzene molecules'',
non-planar nets {\it etc.}, as seen in Figure~\ref{fig:monopoles_strings_Q2}(b).
In the $Q=4$ case the monopole clusters must be similar to fullerene ``cages''
(carbon clusters).

\subsection{Topological interactions}
\label{subsec:Topological_interaction}

Let us  consider now  in more detail
the contribution of the topological defects to the Wilson loop,
$W_{q,Q}$, which defines the potential between two oppositely charged static external sources,
$\langle W_{q,Q}
\rangle \propto \exp\{ - T \, V_{q,Q}(R)\}$. Here we have assumed that
the Wilson loop has a rectangular shape, $T \times R$ with $T \gg R$.
If no matter fields are coupled to the gauge field, $Q=0$,
all non--perturbative contributions to the long-range potential would come from
the monopole fields alone
because the monopole fields are long-ranged in this case.
In the presence of an electrically charged matter field, however,
the magnetic fields are squeezed into vortices, which carry
$\Phi = \frac{2 \pi}{Q}$
units of magnetic flux.
These vortices will provide the major contribution to the potential between
external sources at large distances. The contribution of a general field
configuration to the Wilson loop is
$W^{\cC}_{q,Q} = \exp\{ i~q~\Phi_\cC\}$.
where $\Phi_\cC$ is the Abelian flux going through the contour $\cC$.
Therefore each closed vortex trajectory $j_0$
contributes to the Wilson loop a factor
\beqn
W^{\cC}_{q,Q} = \exp\{ 2 \pi i~\frac{q}{Q}~
\LL(j_0,\cC) \}\,,
\label{eq:linking}
\eeqn
where $\LL(j_0,\cC)$
is the number of vortices piercing
a surface spanned by the Wilson loop contour $\cC$.
The integer number $\LL$ is an analogue of the linking
number between vortex trajectories and the external particle trajectory
discussed below.

The contribution of vortices to the Wilson loop quantum average in the form
\eq{eq:linking} may asymptotically lead to an area law provided the ratio
$q/Q$ is not integer. An example of such a behavior is known from SU(2) gluodynamics
where the so-called center vortices were shown to give a dominant contribution
to the confinement of quarks~\cite{ref:Greensite_Main}.
Thus, we expect that in the $Q>1$ cases the monopoles may form various types of
clusters which range from simple wire-like structures to more complicated
extended three-dimensional structures like nets {\it etc.}

The contribution of such vortices to the external particle potential depends on
the ratio of the external charges, $q$, to the dynamical charges, $Q$.
Eq.~\eq{eq:linking} states that if $\frac{q}{Q} \in \Z$,
then the contribution is trivial while in the case
$\frac{q}{Q} \notin \Z$ the vortices actually contribute to the Wilson loop average.
In the latter case -- as the analogy with SU(2) tells us -- the Wilson loop may
show the area law. Thus we conclude
\beqn
\frac{q}{Q} \in \Z    & \Leftarrow \Rightarrow & \mbox{string breaking}\,,\nonumber\\
\frac{q}{Q} \notin \Z & \Leftarrow \Rightarrow & \mbox{no string breaking}\,.
\label{eq:condition_string_breaking}
\eeqn

It is quite instructive to reproduce these expectations by rewriting
the partition function of the cAHM${}_3$ in the London limit
in terms of topological defects using the so-called
Berezinsky-Kosterlitz-Thouless~\cite{ref:KT,ref:Berezinsky} (BKT)
transformation. For the sake of simplicity we use the Villain-type
form of the action with couplings $\tilde\beta$ and $\tilde\kappa$.
The cAHM${}_3$ partition function in the Villain form has the form
\beqn
\cZ = \int\limits^\pi_{-\pi}  \dD \theta \int\limits^\pi_{-\pi}   \dD \varphi
\sum_{n(c_2)\in\Z}\sum_{l(c_1)\in\Z} \exp\Bigl\{ - \tilde\beta {||\dd \theta + 2 \pi n||}^2
- \tilde \kappa {||\dd \varphi - Q\, \theta + 2 \pi l||}^2\Bigr\}\,,
\label{eq:Z-fields}
\eeqn
where $n$ and $l$ are integer-valued forms
defined, respectively, on plaquettes, $c_2$, and links, $c_1$, of the lattice.
Here and below we use the compact notations of differential forms on the
lattice (for a review see, $e.g.$, Ref.~\cite{Review:Monopole}).
In brief, the plaquette angle $\theta_P$ on the plaquette $\{x,\mu\nu\}$
is written as
${(\dd \theta)}_{x,\mu\nu} = \theta_{x,\mu} + \theta_{x+\hat\mu,\nu} -
\theta_{x+\hat\nu,\mu} - \theta_{x,\nu}$.
The gradient of the phase of the Higgs field
belonging to the link $\{x,\mu\}$ is
${(\dd \varphi)}_{x,\mu} = \varphi_{x+\hat{\mu}} - \varphi_x$.
The Villain couplings are related to the original couplings as follows:
\beqn
  \tilde\beta(\beta) = {\Bigl[ 2 \log
  \frac{I_0(\beta)}{I_1(\beta)}\Bigr]}^{-1}\,,\quad
  \tilde\kappa(\kappa) = {\Bigl[ 2 \log
  \frac{I_0(\kappa)}{I_1(\kappa)}\Bigr]}^{-1}\,.
  \label{eq:BetaV}
\eeqn

Let us apply the BKT transformation~\cite{ref:KT,ref:Berezinsky,Review:Monopole}
first with respect
to the gauge field in Eq.~\eq{eq:Z-fields}.
The sum over the integer rank-two form
$n$ in (\ref{eq:Z-fields}) can be
rewritten\footnote{Here and below we omit volume-dependent constant factors.}
in terms of a sum over an integer-valued one-form $p$ and
an integer-valued three-form $m$,
\beqn
\sum_{n(c_2)\in\Z} = \sum_{p(c_1)\in\Z}
\sum_{m(c_3)\in\Z} \; ,
\label{eq:summation}
\eeqn
using the decomposition
\beqn
n = k[m] + \dd p\,, \quad \mathrm{where} \quad \dd k[m] = m \,.
\label{eq:separation:n}
\eeqn
Here $k[m]$ is an arbitrary integer two-form
satisfying the second equation in~\eq{eq:separation:n}.
On the dual lattice
($i.e.$, on the lattice which is shifted by a half lattice spacing
forward in all three directions)
the form $\dual m$ is an integer-valued scalar field (zero-form)
constrained (due to the second equation in \eq{eq:separation:n})
as follows:
\beqn
\sum_{\dual x} \dual m_{\dual x} = 0 \; ,
\label{eq:neutrality}
\eeqn
what is equivalent to $\dd m=0$ or $\delta \dual m=0$ (with
the co-derivative $\delta= \dual \dd \dual$).

The integer-valued form $\dual m$ represents, if non-vanishing, the
presence of monopoles~\cite{Review:Monopole} on the sites of the
dual lattice. For example, if exactly one monopole (antimonopole)
resides in a particular cube $c_3$ of the original lattice,
then the form $\dual m$ is equal to $\pm 1$
at the corresponding site $\dual x$.
Equation~\eq{eq:neutrality} means that the total magnetic charge
on the lattice is zero.

Using the Hodge-de-Rahm identity
$1 = \delta \Delta^{-1} \dd + \dd \Delta^{-1} \delta$
where $\Delta$ denotes  the lattice Laplacian,
the form $k[m]$ in Eq.~\eq{eq:separation:n} can be rewritten as
$k[m] = \dd \Delta^{-1} \delta k[m] + \delta \Delta^{-1} m$.
Therefore the two-form $n$ can be written as
$n = \delta \Delta^{-1} m + \dd (\Delta^{-1} \delta k[m]+p)$
and we obtain the gauge invariant plaquette in the form
\beqn
\dd \theta + 2 \pi n = \dd \theta^{\mathrm{n.c.}} + 2 \pi \delta \Delta^{-1} m\,,\quad
\theta^{\mathrm{n.c.}} = \theta + 2 \pi \Delta^{-1} \delta k[m] + 2 \pi p\,,
\eeqn
where $\theta^{\mathrm{n.c.}}$ is a non-compact gauge field,
{\it i.e.} $\theta^{\mathrm{n.c.}}_l \in (-\infty,+\infty)$.
Moreover, the sum over the integer-valued variable $p$ and integration
over the compact variable
$\theta$ can be represented (again up to a volume-dependent factor)
as an integration over the non-compact variable $\theta^{\mathrm{n.c.}}$.

Of course, the BKT transformation is nothing but a change of variables.
After the first BKT transformation Eq.~\eq{eq:Z-fields} reads as follows:
\beqn
\cZ \propto \int\limits^\infty_{-\infty}  \dD \theta^{\mathrm{n.c.}} \int\limits^\pi_{-\pi} \dD \varphi
\sum_{\stackrel{\dual m (\dual c_3) \in \Z} {\delta \dual m = 0}}\,
\sum_{l'(c_1)\in\Z} \exp\Bigl\{ - \tilde\beta {||\dd \theta^{\mathrm{n.c.}}||}^2
- 4 \pi^2 \, \tilde\beta (\dual m, \Delta^{-1} \dual m) \nonumber \\
- \tilde \kappa {||\dd \varphi - Q\, \theta^{\mathrm{n.c.}}
+ Q\cdot 2 \pi \Delta^{-1} \delta k[m] + 2 \pi l'||}^2\Bigr\}\,,
\label{eq:Z-fields:2}
\eeqn
where we have used the basic relations of the differential lattice formalism:
$\dd^2 = \delta^2 = 0$, $\Delta = \dd \delta + \delta \dd$, $\dual \dd \dual = \delta$, and
$*^2 = 1$.
In addition, the original one-form $l$ is replaced through a shift by $l'=l + Q \, p$
to make the link-dependent part of the action (proportional to $\tilde\kappa$) independent of $p$.
The global condition~\eq{eq:neutrality} is implicitly understood in
Eq.~\eq{eq:Z-fields:2} and subsequent equations.

Next, we apply the BKT transformation
a second time, with respect to the compact phase of the Higgs field. We proceed
similarly to the case of the compact gauge field,
representing the sum over the one-form $l'$ as
sums over some two-form $j$ and a zero-form $r$.
The forms $j$ and $r$ are related to $l'$ as follows:
\beqn
l' = s[j] + \dd r\,, \quad \mathrm{where} \quad \dd s[j] =  j\,.
\label{eq:separation:l}
\eeqn
Again using the Hodge-de-Rahm identity we introduce the non-compact scalar field $\varphi^{\mathrm{n.c.}}$:
\beqn
\dd \varphi + 2 \pi l' = \dd \varphi^{\mathrm{n.c.}} + 2 \pi \delta \Delta^{-1} j\,,\quad
\varphi^{\mathrm{n.c.}} = \varphi + 2 \pi \Delta^{-1} \delta s[j] + 2 \pi r\,.
\eeqn
The partition function~\eq{eq:Z-fields:2} is now written as follows:
\beqn
\cZ \propto \int\limits^\infty_{-\infty}  \dD \theta^{\mathrm{n.c.}}
\int\limits^\infty_{-\infty} \dD \varphi^{\mathrm{n.c.}}
\sum_{\stackrel{\dual m (\dual c_3) \in \Z} {\delta \dual m = 0}}\,
\sum_{\stackrel{\dual j(\dual c_2)\in\Z}{\delta \dual j = 0}}
\exp\Bigl\{ - \tilde\beta {||\dd \theta^{\mathrm{n.c.}}||}^2
- 4 \pi^2 \, \tilde\beta (\dual m, \Delta^{-1} \dual m) \nonumber \\
- \tilde \kappa {||\dd \varphi^{\mathrm{n.c.}} - Q\, \theta^{\mathrm{n.c.}}
+ 2 \pi \Delta^{-1} \delta (j + Q \, k[m])||}^2\Bigr\}\,.
\label{eq:Z-fields:3}
\eeqn
The integer-valued one-form
\beqn
\dual j_m =  Q \dual k[m] + \dual j
\label{eq:modification}
\eeqn
represents, if non-vanishing, the vortex
line defined on a dual link, similarly to the
monopole zero-form $\dual m$.
The constraint
\beqn
\delta \dual j_m = Q \dual m
\label{eq:constraintjm}
\eeqn
(whereas $\delta \dual j = 0$)
indicates conservation of magnetic flux: vortices carry a fraction
${1}/{Q}$ of the total magnetic flux emanating from a monopole or absorbing by
an anti-monopole.

It is self-evident that the definitions, introduced in passing,
of monopoles $\dual m$ and vortices $\dual j_m$ are identical to those
obtained directly from the gauge invariant plaquette
$P= \dd \theta + 2 \pi n$
\beqn
m = \frac{ \dd P}{2 \pi}= \dd n  \,, \quad \dd m =0
\label{eq:mdef}
\eeqn
and  the link $L= \dd \varphi - Q \, \theta + 2 \pi l$
\beqn
j_m= \frac{\dd L + Q\, P}{2\pi}=\dd l + Q \,n \,, \quad \dd j_m= Q \,m \,.
\label{eq:jdef}
\eeqn

Performing the Gaussian integration over the non-compact fields~\footnote{This can be
done in the simplest way by gauge-fixing the Higgs field angles
such that $\dd \varphi^{\mathrm{n.c.}}=0$.}
in Eq.~\eq{eq:Z-fields:3}
we rewrite the partition function in terms
of monopoles and vortices:
\beqn
\cZ =
\sum_{\stackrel{\dual m (\dual c_3) \in \Z} {\delta \dual m = 0}}\,
\sum_{\stackrel{\dual j_m(\dual c_2)\in \Z}{\delta \dual j_m = Q  \dual m}}
\exp\Bigl\{- S_d(\dual m,\dual j_m)\Bigr\}\,.
\label{eq:Z-objects}
\eeqn
The action of the topological defects is
\beqn
S_d(\dual m,\dual j_m) = 4 \pi^2 \tilde\beta (\dual m, \frac{1}{\Delta + M^2} \dual m)
+ 4 \pi^2 \tilde\kappa (\dual j_m, \frac{1}{\Delta + M^2} \dual j_m)\,,
\label{eq:S-objects}
\eeqn
where $M = Q {(\tilde\kappa/\tilde\beta)}^{1/2}$ is the tree-level mass of the
gauge boson.

Let us now consider the contribution of the vortices and monopoles to the potential
$V_{q,Q}(R)$ between a pair of external particles with charges $\pm q$, separated by
a distance $R$.
The potential is given by the quantum average of the Wilson loop defined as
\beqn
      \langle W_q (J) \rangle_Q &=&
      \frac{1}{\cZ}
\int\limits_{-\pi}^{\pi}\dD \theta\int\limits_{-\pi}^{\pi}\dD \varphi
   \sum_{n(c_2)\in \Z} \, \sum_{l(c_1)\in \Z} \times
   \nonumber
   \\
 &&
   \exp\Bigl\{-\tilde\beta || \dd \theta + 2 \pi n||^2-
            \tilde\kappa ||\dd \varphi - Q \theta + 2 \pi l ||^2
            + i q (J,\theta)\Bigr\}
            \,.
\eeqn
The external current satisfies $\delta J$=0.

Following the transformations which led
us from Eq.~\eq{eq:Z-fields} to Eq.~\eq{eq:Z-objects}, the {\it v.e.v.} of the Wilson
loop can be rewritten as
\beqn
\langle W_q \rangle_Q = \langle W_q \rangle_{\mathrm{ph}} \,
                        \langle W_q \rangle_{d} \, .
\label{eq:factorization}
\eeqn
Here
\beqn
\langle W_q \rangle_{\mathrm{ph}} \propto \exp\Bigl\{ - \frac{q^2}{4 \tilde\beta}
\, (J,\frac{1}{\Delta + M^2} J)\Bigr\}
\label{eq:massive_photon_part}
\eeqn
is the perturbative self-interaction contribution of the external loop
given by exchange of a massive photon,
while the non-perturbative factor is due to the topological defects:
\beqn
\langle W_q \rangle_{d} = \frac{1}{\cZ_d}
\sum_{\stackrel{\dual m (\dual c_3) \in \Z} {\delta \dual m = 0}}\,
\sum_{\stackrel{\dual j_m(\dual c_2)\in \Z}{\delta \dual j_m = Q  \dual m}}
\exp\Bigl\{- S_d(\dual m,\dual j_m) - S_{\mathrm{int}}(\dual m,\dual j_m;J)
\Bigr\}
\label{eq:Wd}
\eeqn
with
\beqn
S_{\mathrm{int}} = - 2 \pi i \frac{q}{Q} (\delta j_m, \frac{1}{\Delta + M^2} J)
+ 2 \pi i \frac{q}{Q} \LL(\dual j_0,J) \,.
\label{eq:Sint}
\eeqn
The first term in $S_{\mathrm{int}}$ consists of a Yukawa-type interaction
between the vortices and the external charged particle.
Since the vortices and
the monopoles are related to each other by the constraint~\eq{eq:constraintjm}, the
first term in $S_{\mathrm{int}}$ also represents an indirect interaction between the
monopoles and the external charged particle.

The second term is of topological nature. It is given by the linking number
\beqn
\LL(\dual j_0,J) = (\delta j_0, \Delta^{-1} J) \in \Z
\label{eq:linkingformula}
\eeqn
between the external particle trajectory $J$ and
the one-form $\dual j_0$. The form $\dual j_0$ is derived from the vortex
ensemble $\dual j_m$ by subtracting $Q \dual k[m]$,
{\it cf.} (\ref{eq:modification}),
such that $\delta \dual j_0 = 0$. In contrast to $\dual j_m$, the modified
vortex ensemble $\dual j_0$ consists of oriented and closed loops.
The linking number between $\dual j_0$ and $J$ corresponds to the Aharonov-Bohm
(AB) effect~\cite{ref:MIP_Zubkov}.

In the cQED-like limit, $M \to 0$, the interaction term \eq{eq:Sint}
reduces to the usual cQED-like interaction between monopoles:
\beqn
\lim\limits_{M \to 0} S_{\mathrm{int}} & = &
- 2 \pi i \frac{q}{Q} (\delta j_m, \Delta^{-1} J)
+ 2 \pi i \frac{q}{Q} \LL(\dual j_0,J)  \nonumber\\
& = & 2 \pi i \frac{q}{Q}
(\delta (j_0 - j_m), \Delta^{-1} J) = - 2 \pi i  (k[m], \Delta^{-1} J) \,,
\label{eq:Sint:cQED}
\eeqn
where we have used Eq.~\eq{eq:linkingformula} and Eq.~\eq{eq:modification}.
The interaction~\eq{eq:Sint:cQED} does not depend on the shape of the Dirac
line, $\dual k[m]$, and does depend on the monopole
magnetic charges and monopole locations, $\dual m \equiv \delta \dual k[m]$.

Note that the separation of a general vortex ensemble,
$\dual j_m = \dual j_0 + \dual j_m'$, into closed vortices,
$\delta \dual j_0 = 0$, and open ones, $\delta \dual j_m' = \dual m$, is ambiguous.
In the sum~\eq{eq:Wd} the ambiguity disappears since all ambiguous separations
enter Eq.~\eq{eq:Wd} with the same weight.

The contribution of vortices to the inter-particle potential $V_{q,Q}(R)$
is twofold. First, the closed vortices
$\dual j_0$ interact with the electrically charged particles
via the AB effect if the ratio $q/Q$ is {\it not integer}.
Secondly, the vortices influence the monopole dynamics
via the constraint
(\ref{eq:constraintjm}), contributing in this way to the potential
indirectly, via the monopoles and their interaction with the test
particles.

Actually, orientable vortices with closed magnetic flux lines cannot lead
to confinement of electric charges~\cite{ref:GubarevAHM}.
In that paper numerical studies for the {\it non-compact} version of the
AHM${}_3$ have shown that confinement is not realized.
This is a model in which {\it orientable} vortices are present (and
consequently, an AB effect might be at work) whereas magnetic monopoles
are absent. On the other hand, a long-range interaction between
vortices and the electric Wilson loop may be provided
only by the AB effect in a form visible in Eq.~\eq{eq:Wd}.
In the model presently under consideration the linking of
the {\it non-orientable} vortices with the external electric current is
{\it formally expressed} by (\ref{eq:linkingformula}) in terms of the
{\it redefined} closed and orientable loops $\dual j_0$.
Making this point, we are implicitly leaning upon that we are working in
a model with a mass gap in which any interactions -- except for topological
ones like the one mediated by the linking number -- must be suppressed.

In four dimensional QCD the so-called center vortices are known to lead to the
area law of the Wilson loop via the AB interaction~\cite{ref:Greensite_Main}.
They are providing a dominant contribution to the confinement of quarks if they
are localized by means the so-called Maximal Center projection.
There is no contradiction between the roles of the AB effect in QCD
and the non-compact AHM since in the QCD case the center vortices are
non-orientable worldsheets whereas in the case of the 3D non-cAHM they are
orientable lines.
Moreover, considered in an Abelian sense, the non-orientable center vortices
can be represented as piecewise orientable vortices with monopole trajectories
separating segments with different flux orientations from each other.
The required non-orientability implies the existence of monopoles living
on the vortex sheets.

We will see that all percolating loops in the case of the model studied
here (which are responsible for confinement) really form an alternating
monopole-antimonopole chain which makes them {\it physically non-orientable}.
Thus, we expect that in cAHM${}_3$ the vortices play a role to accomplish
both confinement and string breaking phenomena, for the respective sources,
by restructuring the monopoles configurations.

\section{Phase structure in the London limit and the percolation properties
of monopole chains}
\label{sec:phase_structure_percolation}

\subsection{The finite-temperature case}
\label{subsec:finite_temperature_case}

In this subsection
we report on our numerical study of the confining properties
as seen by static external particles with different electric charges $q$
in the case of the cAHM${}_3$ with doubly--charged matter.
In the confining phase of this model
only external particles with charge $q=1$ must be
asymptotically confined while particles with charge $q=2$ should experience
the flattening of the potential at large distances.

To observe this effect we have simulated the model~\eq{eq:London-action} first
on lattices $32^2\times 8$.
The choice of the asymmetric lattice
(finite $T$) was not a matter of principle but
simply dictated by the fact that in the
case of symmetric ($L^3$) lattices the potential can be measured only using
Wilson loops which are notoriously unsuitable for an
observation of the string breaking effect. We have performed
simulations at fixed gauge coupling constant, $\beta=1.2$, the choice
of which was motivated by visualization reasons. Indeed, in order to
clearly see the monopole structures, the density of the monopoles must
be neither too high nor too low. According to Eq.~\eq{eq:S-objects} a
singly--charged (anti)monopole enters the partition
function with the action $S^{(1)}_m = 4 \pi^2 \tilde \beta \,
\Delta^{-1}(0)$ (with $\Delta^{-1}(0) \approx 0.25$ in the thermodynamic limit).
Thus an increase (decrease) of the gauge coupling $\beta$
constant suppresses (enhances) the number of monopoles. The chosen
value of $\beta$ is ideal from the point of view of our purposes. We have
used about $10^5$ configurations for each value of the hopping
parameter $\kappa$. Some results on the finite--temperature
case (partially included in this subsection) were previously reported in
Ref.~\cite{ref:U1Q2PLB}.

\begin{figure*}[!htb]
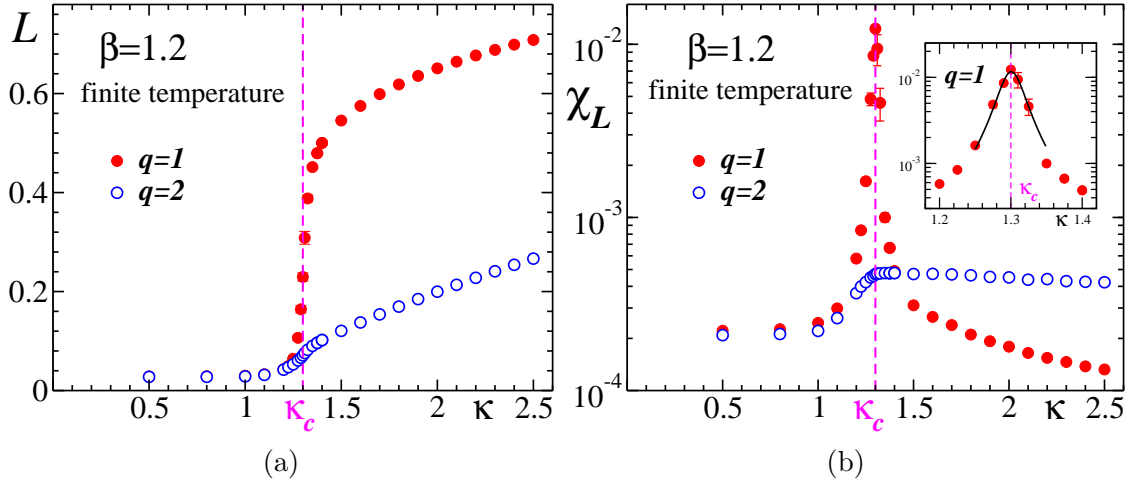

\begin{center}
\begin{tabular}{cc}
\includegraphics[scale=0.34,clip=true]{btpolyakov.loop.eps} &
\includegraphics[scale=0.34,clip=true]{btpolyakov.susc.eps} \\
(a) & (b) \\
\end{tabular}
\end{center}
\vskip -5mm
\caption{The $q=1,2$ Polyakov loops (a) and their susceptibilities {\it vs.} $\kappa$
at $\beta=1.2$ on a $32^2 \times 8$ lattice. The insert in Figure (b) shows the
behavior of the $q=1$ susceptibility in the vicinity of the phase transition
with the fit according to Eq.~\eq{eq:fitting_function}.}
\label{fig:polyakov_loops}
\end{figure*}

To probe the phase structure of the model we first
study the vacuum expectation values ({\it v.e.v.}'s)
of the $q=1$ and $q=2$ Polyakov loops,
\beqn
L_q(\vec x) = \exp\{ i q \sum_{t=1}^{L_t} \theta_{4}(\vec x, t)\}\,.
\eeqn
Their expectation values are shown in Figure~\ref{fig:polyakov_loops}(a)
as a function of the hopping parameter $\kappa$. At small (large) $\kappa$
the $v.e.v.$'s of both loops are low (high) corresponding to the confinement (Higgs)
phase. One can clearly see that the (outside the Higgs phase) confined $q=1$
external charges more sensitively react to the onset of the Higgs
phase compared to the $q=2$ charges. This is a typical
signature of the $q=2$ string breaking.
The abundantly available $Q=2$ particles screen $q=2$ external charges
which leads to the low sensitivity of the $q=2$ loops to the phase transition.

In Figure~\ref{fig:polyakov_loops}(b) we show the Polyakov loop susceptibilities
{\it vs.} $\kappa$. The $q=2$ susceptibility does not indicate any significant change
at the transition point while the $q=1$ susceptibility shows a sharp peak
signalling the presence of a phase transition.
We have fitted the $q=1$ susceptibility
in the vicinity of the maximum by the function~\cite{Chernodub:2002ym}:
\beqn
\chi^{\mathrm{fit}}(\kappa)
= \frac{C_1}{{(C_2 +{(\kappa - \kappa_c)}^2)}^{\gamma_\chi}}\,,
\label{eq:fitting_function}
\eeqn
with $C_{1,2}$,
$\gamma_\chi$
and $\kappa_c$ being the fit parameters. From the fit
(shown in the insert of Figure~\ref{fig:polyakov_loops}(b) as a solid line)
we obtain that the transition happens at
\beqn
\kappa_c\equiv\kappa^{q=1}_c = 1.300(1)\,.
\eeqn
To fit the susceptibility in the vicinity of the critical value we have chosen
the region such that $\chi^2/d.o.f. \sim 1$. This choice applies to all fits of
susceptibilities performed in this paper.

\begin{figure*}[!htb]
\begin{center}
\includegraphics[scale=0.45,clip=true]{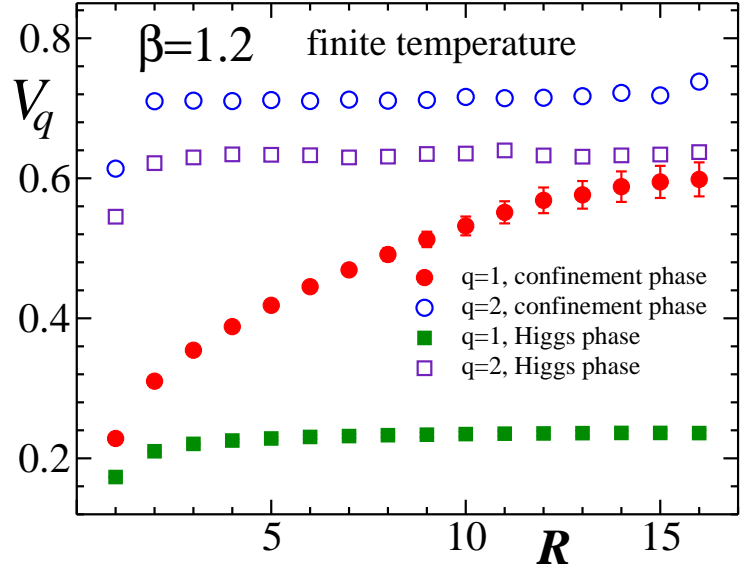}
\end{center}
\vskip -5mm
\caption{The $q=1,2$ potentials in the confinement ($\kappa = 1.275$)
and the Higgs ($\kappa = 1.325$) phases at $\beta=1.2$ (from
Ref.~\cite{ref:U1Q2PLB}).}
\label{fig:fromPLB}
\end{figure*}

The static potentials between $q$--charged external particles can be expressed via
the Polyakov loop correlators,
$V_{q,Q}(R) = - L^{-1}_t\, \log \langle L_q({\vec 0}) L^*_q({\vec R})\rangle$.
As it was shown in Ref.~\cite{ref:U1Q2PLB}, in the confinement phase the potential
for the $q=1$ external charges is linearly rising at large distances while $q=2$
potential shows a rapid flattening corresponding to dynamical particles
popping up from the vacuum to cause string breaking. In the Higgs phase {\it all} potentials
suffer flattening due to deconfining nature of this phase.
The corresponding potentials are shown in Figure~\ref{fig:fromPLB}.
This result completely
agrees with the general form of the phase diagram shown in
Figure~\ref{fig:phase_diagram_general}.

As we have mentioned above, the cAHM contains
two types of topological objects -- monopoles and vortices.
The simplest characteristic of a topological defect is its density.
The monopole and the vortex densities are (the volume is $V= L^2 L_t$ for
finite temperature and $V=L^3$ at $T=0$)
\beqn
  \rho_{\mathrm {mon}} = \frac{1}{V} \sum_{\dual c_3} |\dual m| \,,\qquad
  \rho_{\mathrm {vort}} = \frac{1}{3 V} \sum_{\dual c_2} |\dual j_m| \,,
\eeqn
respectively. The monopole charge is defined in the standard way
(see also (\ref{eq:mdef})),
\beqn
  m = \frac{1}{2 \pi} \dd {[\dd \theta]}_{2\pi} \,,
\eeqn
where
${[\cdots]}_{2\pi}/(2\pi)$ denotes the integer part modulo $2 \pi$.
Following Ref.~\cite{ANO_definition}, the vortex current is defined as
(cf. (\ref{eq:jdef}))
\beqn
j = \frac{1}{2 \pi} (\dd {[\dd \varphi + \, \theta]}_{2\pi} - Q
{[\dd \theta]}_{2\pi}) \,.
\eeqn

\begin{figure*}[!htb]
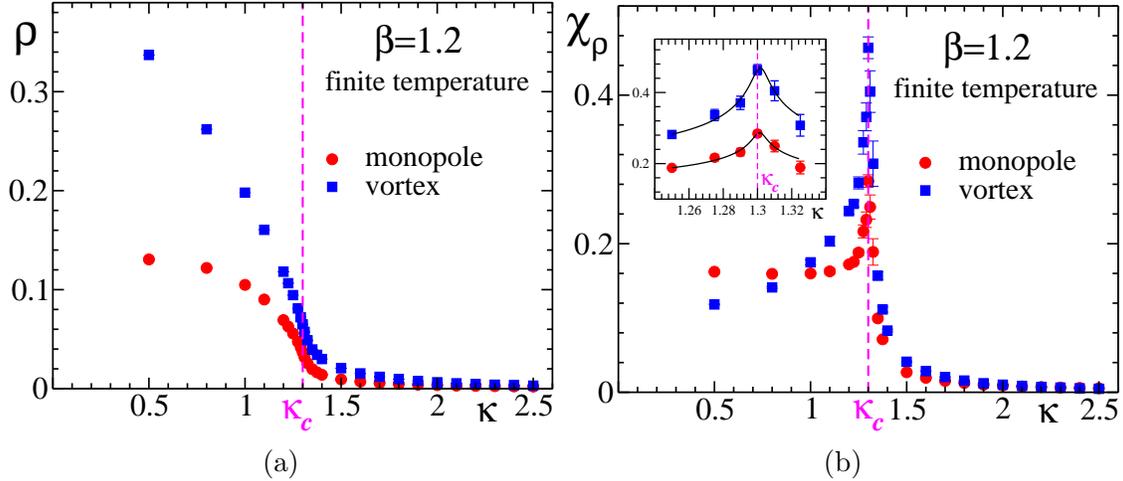

\begin{center}
\begin{tabular}{cc}
\includegraphics[scale=0.34,clip=true]{btdensities.eps} &
\includegraphics[scale=0.34,clip=true]{btdensity.susc.eps}\\
(a) & (b)
\end{tabular}
\end{center}
\vskip -5mm
\caption{(a) The monopole and vortex densities (from Ref.\cite{ref:U1Q2PLB}) and (b)
their susceptibilities, $vs.$ $\kappa$ at $\beta=1.2$ on $32^2\times 8$ lattice.
The insert in Figure (b) shows the behavior of the susceptibilities in the vicinity
of the phase transition. The solid lines show the fits by
Eq.~\eq{eq:fitting_function}.}
\label{fig:monopole_defects}
\end{figure*}

According to Figure~\ref{fig:monopole_defects}(a)
the densities (exposed in lattice units)
of both monopoles and vortices are gradually
decreasing functions of the coupling $\kappa$
what is partly a finite temperature effect.
In the confinement (Higgs) phase
the density of both vortices and monopoles is high (low),
in agreement with the expectation that the confining properties of
the cAHM${}_3$ are due to
topological defects.
Both susceptibilities are peaked at the phase transition point.
The fits by Eq.~\eq{eq:fitting_function} of the monopole and vortex susceptibilities
give the following transition values of the hopping parameter, respectively:
\beqn
\kappa^{\mathrm{mon}}_c=1.301(3)\,,\qquad \kappa^{\mathrm{vort}}_c=1.302(2)\,.
\eeqn
These values are consistent with each other as well as with the critical value
obtained with the help of the $q=1$ Polyakov loop susceptibility discussed above.

Obviously, in the confining phase the monopoles cannot form
a monopole or dipole plasma because in these cases both external charges
$q=1,2$ would be simultaneously confined or deconfined.
Thus the only possible kind of monopole configurations --
which can explain both the linearly rising potential for the $q=1$ electric
charges and the string breaking for the $q=2$ charges -- is a monopole chain
schematically plotted in Figure~\ref{fig:monopoles_strings_Q2}(a).
In a monopole chain the monopoles and antimonopoles are mutually alternating.
Thus the magnetic flux coming from a monopole inside the chain is separated
into two parts, squeezed into the vortices and, as a consequence,
forming a {\it non-orientable} closed magnetic flux. Each piece of such a vortex
carries (in average) a half flux, $\Phi = \frac{2 \pi}{Q} \equiv \pi$. If such a
flux pierces the $q$-charged Wilson loop it provides a (multiplicative) contribution
to the loop close to ${(-1)}^q$. This leads to a disorder for odd charged external
particles (eventually leading to a linearly rising potential) whereas particles
with even charge are not confined. The described picture is very similar to what is
observed~\cite{ref:Greensite_correlations} in QCD for the correlations between the
Abelian monopoles and the center vortices.

\begin{figure*}[!htb]
\begin{center}
\begin{tabular}{cc}
\includegraphics[scale=0.65,clip=true]{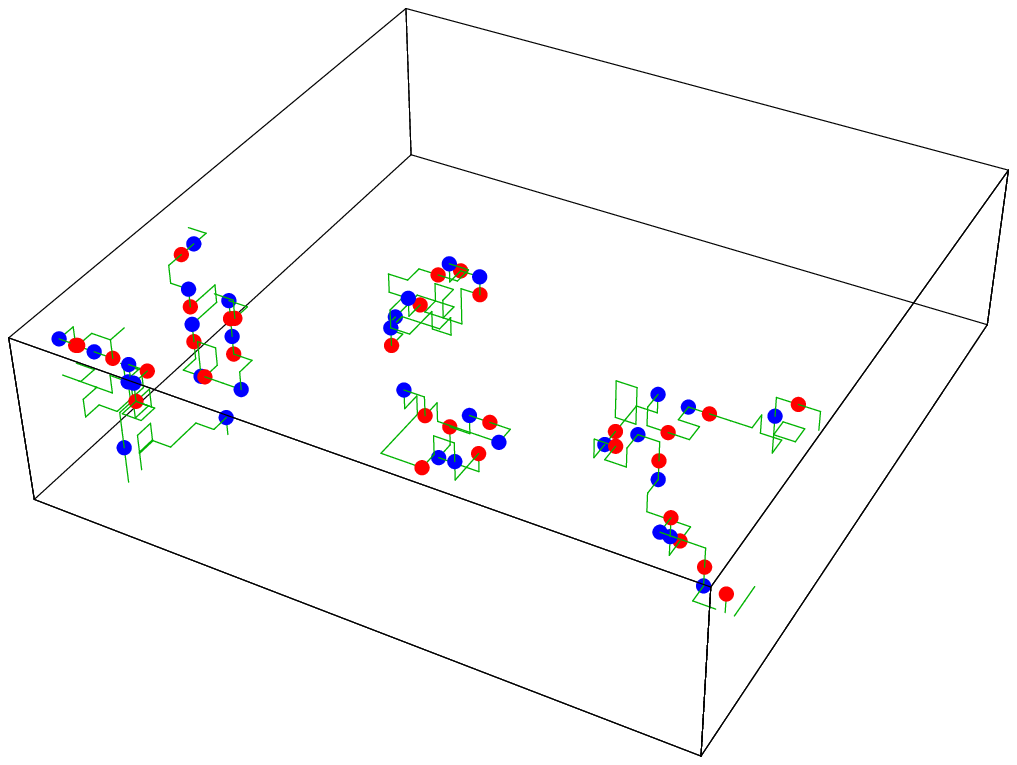} &
\includegraphics[scale=0.65,clip=true]{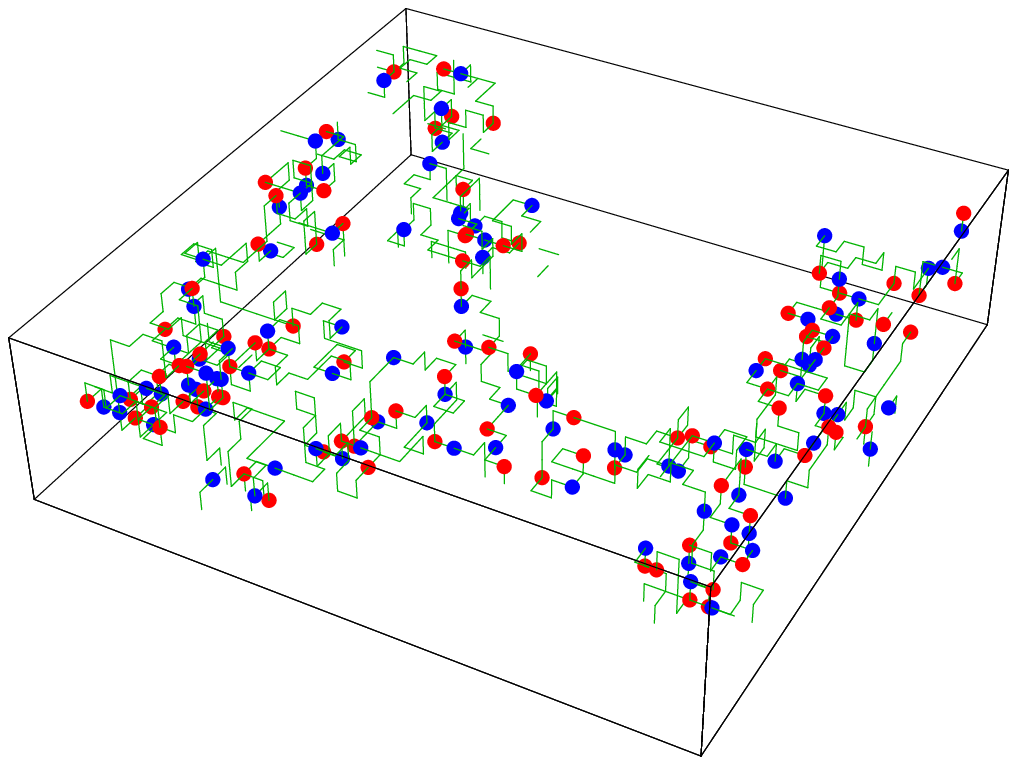} \\
  (a) & \hspace{5mm} (b)  \\[3mm]
\includegraphics[scale=0.65,clip=true]{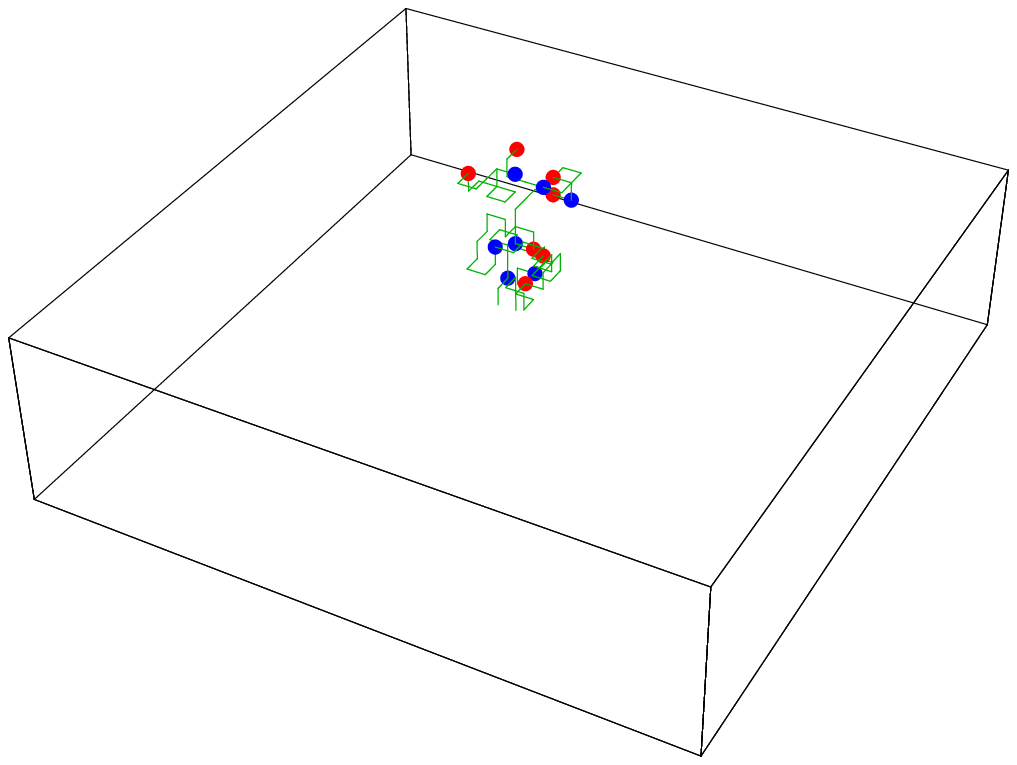} &
\includegraphics[scale=0.65,clip=true]{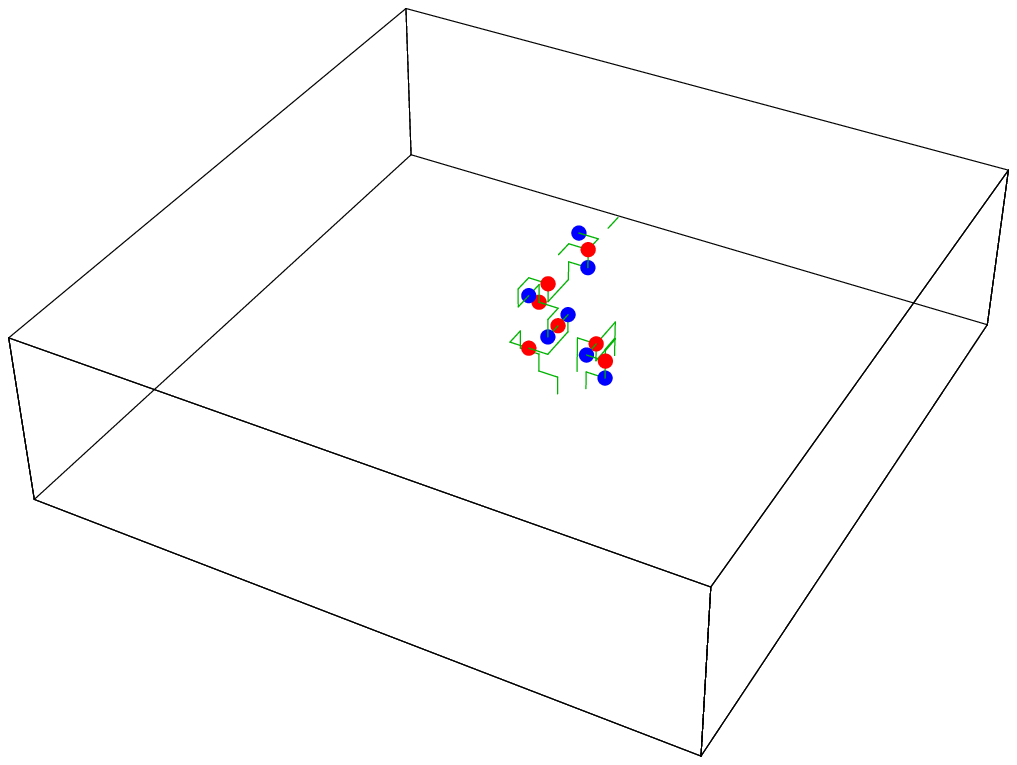} \\
  (c) & \hspace{5mm} (d)
\end{tabular}
\end{center}
\vskip -5mm
\caption{Examples of the monopole configurations in the confinement phase
at $\kappa=1.275$ (upper row) and in the deconfinement (Higgs) phase
at $\kappa=1.325$ (lower row). The (anti-)monopoles are shown by the dots
while the vortices are shown by the solid lines.}
\label{fig:examples}
\end{figure*}

In Figures~\ref{fig:examples}(a-d)
we show typical monopole configurations observed in our numerical simulations.
In the confining phase the monopoles form chains which either wrap around
the temperature direction (Figure~\ref{fig:examples}(a)) or fill all the lattice
volume (Figure~\ref{fig:examples}(b)). In the last case the monopole chains are said
to be percolating. In the Higgs phase the monopole chains wrap only around
the temperature direction (Figure~\ref{fig:examples}(c,d)).
In the next subsection we investigate the percolation properties of the monopole
chains in the zero--temperature model. We want to demonstrate
that the mechanism causing the confining/deconfining effects is the same as for
the finite temperature lattice dealt with in this subsection.

\subsection{The zero-temperature case at fixed gauge coupling $\beta=2.0$}
\label{subsec:T0_at_beta2.0}

We report now the study of the doubly--charged cAHM
at zero-temperature, on the lattice $16^3$.
{}From $2\cdot 10^3$ to $8\cdot10^4$ measurements for
every $\beta,\kappa$ pair have been collected.
First we map out the phase structure along a vertical line at $\beta=2.0$ in
the phase diagram Fig. \ref{fig:phase_diagram_general}, varying the hopping
parameter $\kappa$. We mainly concentrate on the quantitative analysis of
the monopole chains and their components, vortices and monopoles.

\begin{figure*}[!htb]
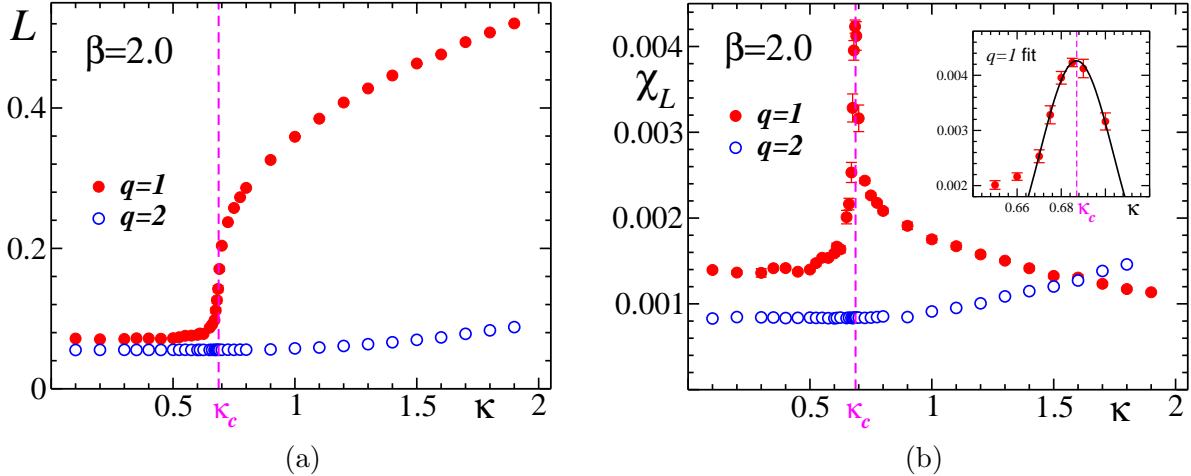

\begin{center}
\begin{tabular}{cc}
\includegraphics[scale=0.34,clip=true]{bpolyakov.loop.eps} \hspace{5mm} &
\includegraphics[scale=0.34,clip=true]{bpolyakov.susc.eps} \\
  (a) & \hspace{5mm} (b)  \\
\end{tabular}
\end{center}
\vskip -5mm
\caption{(a) The expectation values of the Polyakov loop for $q=1$ and $q=2$
and (b) their susceptibilities {\it vs.} $\kappa$ for a lattice $16^3$.
The insert in Figure (b) shows the behavior of the $q=1$ susceptibility in
the vicinity of the phase transition
and the fit according to Eq.~\eq{eq:fitting_function}.}
\label{fig:bpolyakov_susc}
\end{figure*}

Similarly to the finite--temperature case, for orientational purposes,
we measure the expectation values of
the Polyakov loops which also in the zero--temperature model
is defined as a Wilson line closed via the boundary conditions.
The expectation values of the $q=1,2$ Polyakov loops are shown in
Figure~\ref{fig:bpolyakov_susc}(a).
As in the finite $T$ case the {\it v.e.v.} of the $q=1$ Polyakov loop
is noticeably larger in the Higgs phase (larger $\kappa$'s) and
almost constant in the confinement phase (smaller $\kappa$'s). The
$q=2$ Polyakov loop is almost constant crossing the transition.
Consequently, the $q=1$ susceptibility of the Polyakov loop is peaked at the transition point
for moderate lattice volume while
the susceptibility of the $q=2$ line is insensitive to the transition. The fit of the
$q=1$ susceptibility by Eq.~\eq{eq:fitting_function} is shown in the insert of
Figure~\ref{fig:bpolyakov_susc}(b)
as a solid line. The corresponding critical value of the hopping parameter is
\beqn
\kappa_c\equiv \kappa^{q=1}_c = 0.687(1)\,.
\eeqn

\begin{figure*}[!htb]
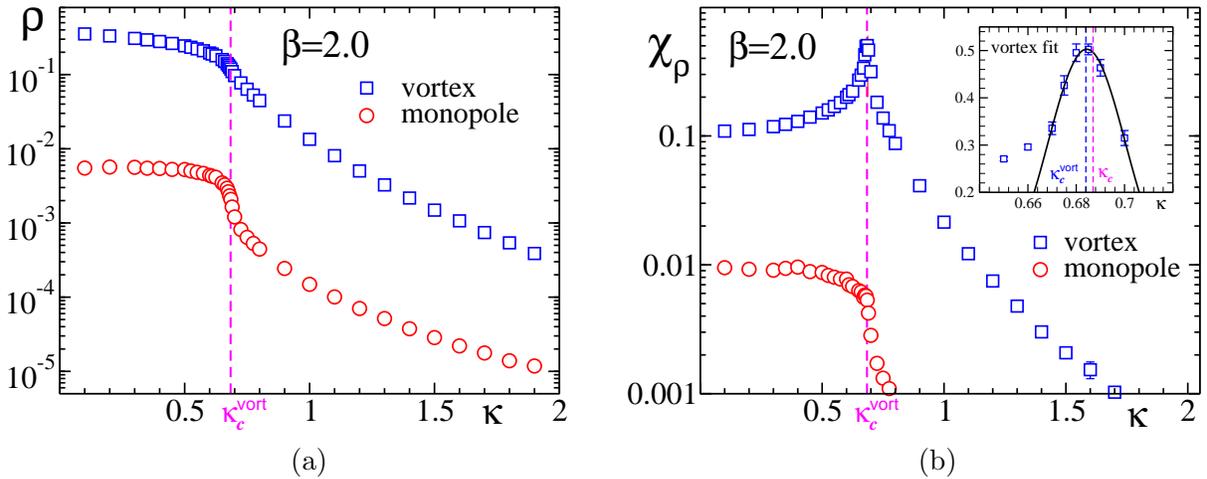

\begin{center}
\begin{tabular}{cc}
\includegraphics[scale=0.34,clip=true]{bdefect.dens.eps} \hspace{5mm} &
\includegraphics[scale=0.34,clip=true]{bdefect.susc.eps} \\
  (a) & \hspace{5mm} (b)
\end{tabular}
\end{center}
\vskip -5mm
\caption{(a) The densities of the topological defects and (b) their
susceptibilities {\it vs.} $\kappa$. The insert in Figure (b) shows
the behavior of the vortex susceptibility in the vicinity
of the phase transition. The solid line shows the fit by
Eq.~\eq{eq:fitting_function}.}
\label{fig:bdefect_susc}
\end{figure*}

To clarify the role of the topological defects we
plot the density of defects in Figure~\ref{fig:bdefect_susc}(a).
One can clearly see that
in this region of the phase diagram the density of monopoles is much smaller
than the density of vortices. This is not unexpected since at $\beta=2.0$
the action of an isolated monopole is
$S^{(1)}_m = 4 \pi^2 \tilde \beta \, \Delta^{-1}(0) \approx 20$, therefore the
monopoles must be suppressed drastically. One may therefore
conjecture that at $\beta=2.0$ the leading role at the transition should be
played by vortices and not by the monopoles, in other words,
the dynamics of the monopoles is driven by the vortex networks.

In Figure~\ref{fig:bdefect_susc}(b) we show the susceptibilities of the monopole and vortex densities. The vortices
are sensitive to the phase transition while the
susceptibility of the
monopole density does not show any noticeable peak at the
critical value of the hopping parameter. The fit of the susceptibility of the vortex density by
function~\eq{eq:fitting_function} gives the pseudo-critical value of the hopping parameter
\beqn
\kappa^{\mathrm{vort}}_c = 0.684(3)\,,
\eeqn
consistent with the value obtained from the susceptibility
of the Polyakov loop.

\begin{figure*}[!htb]
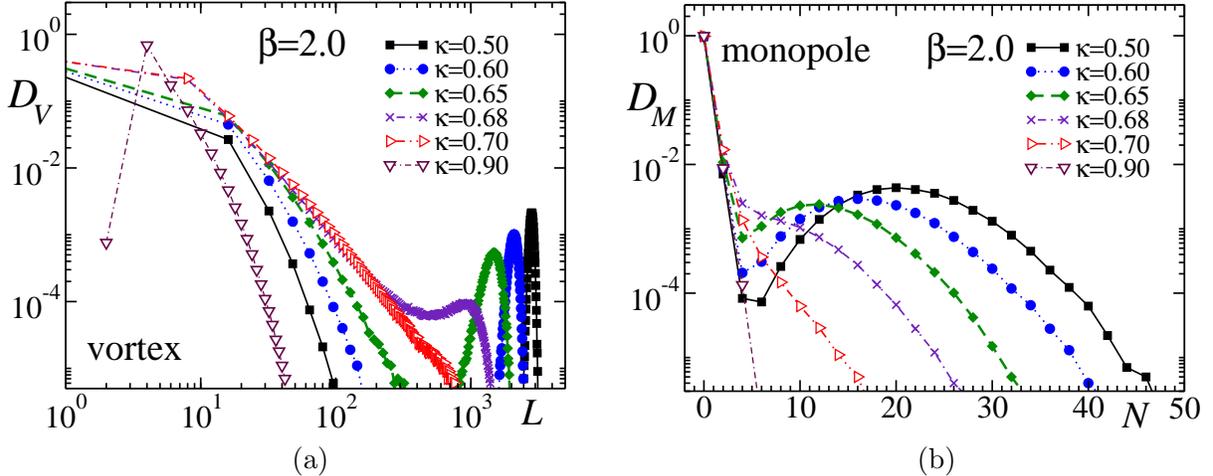

\begin{center}
\begin{tabular}{cc}
\includegraphics[scale=0.34,clip=true]{bvorthist.eps} \hspace{5mm} &
\includegraphics[scale=0.34,clip=true]{bmonhist.eps} \\
  (a) & \hspace{5mm} (b)
\end{tabular}
\end{center}
\vskip -5mm
\caption{Histograms of the (a) vortex and (b) monopole distributions.}
\label{fig:clusters_beta}
\end{figure*}

To characterize quantitatively the behavior of the clusters we plot in
Figure~\ref{fig:clusters_beta}(a)
the distribution $D_V$ of the lengths $L$
of the vortices inside the (mutually disconnected) clusters. The distributions
for various $\kappa$ are normalized to unity. To see the difference between
the phases, we plot the distributions for $\kappa$ values in the vicinity
of the phase transition. One can see that in the confinement phase the
vacuum is populated by two different types of clusters since the distributions
have two peaks. The vortex lengths in the first type of clusters
is typically short. The vortices of this type are called ultraviolet vortices.
The term ``ultraviolet'' is borrowed from $4D$ QCD which also possesses Abelian
monopole trajectories (defined in an Abelian projection) which fall into
similar classes of loops. In QCD the ultraviolet monopoles can be described
by a formula $D_M^{UV}
\sim L^{-\lambda}$ where $\lambda$
is a positive number which does not depend on the lattice
volume~\cite{ref:clusters}. This experience leads us to
conjecture a similar behavior in the $3D$ cAHM as well, except that the
loops are now formed by vortices.

The other population of vortices
consists of the so called ``infrared'' vortices which form the second peak
in the distribution $D_V$. The average number of the length $L$ of the
vortices in this peak is proportional to the number of links
on the lattice, $3 V$, multiplied by an ``infrared density'' $\rho^{V}_{IR}$
which may be also referred to as the ``vortex condensate''. The vortices from
this condensate occupy uniformly all the volume of
the system and their typical length is equal to $\rho^{V}_{IR} \, 3 V$.
The fluctuations of the length of the infrared clusters
are expected to be random. Therefore, they may be described by a Gaussian ansatz
(here we refer to Ref.~\cite{ref:gaussian_QCD} where a similar observation
was made for monopole trajectories).
It is clear that the vortex condensate exists only in the confinement phase
(signalled by the infrared peak of the histogram) while
it is absent in the Higgs (deconfinement) phase

A similar behavior is also observed for the distributions
$D_M$ with respect to the monopole number $N$ inside the mutually disconnected
vortex clusters. Monopoles are pointlike and connected by vortices\footnote{Inspecting the
vortex clusters {\it without} monopole-antimonopole pairs
in the confinement phase we notice that they
belong to the ultraviolet component (loops of short length)
which is irrelevant for topological (AB) confinement.}.
These distributions are plotted in Figure~\ref{fig:clusters_beta}(b). One can
also see the two-peak structure in the confinement phase
(the infrared and the ultraviolet monopoles corresponding to long and short
vortex clusters) and the disappearance of the infrared peak in
the Higgs (deconfinement) phase.

{}From Figures~\ref{fig:clusters_beta}(a,b) we
conclude that in the confinement phase a vortex condensate is formed. The
percolating vortices are carrying the monopoles which are responsible for the
confining properties of the vacuum.

In order to characterize the cluster distributions of vortices and monopoles
$D_{V,M}$ we fitted the distribution functions by
\beqn
D^{\mathrm{fit}}_V(L) = C_{1,V}\, L^{-\lambda_V} + C_{2,V} \,
\exp\Bigl\{ - \alpha_{V} (L - \rho^{V}_{IR} \cdot 3 V)^2 \Bigr\}
\label{eq:fit_clusterV}
\eeqn
and by an analogous form for the monopole distributions
\beqn
D^{\mathrm{fit}}_M(N) = C_{1,M}\, N^{-\lambda_M} + C_{2,M} \,
\exp\Bigl\{ - \alpha_{M} (N - \rho^{M}_{IR} \cdot V)^2 \Bigr\}
\label{eq:fit_clusterM}
\eeqn
to determine the condensates $\rho^{V,M}_{IR}$.
Equations~\eq{eq:fit_clusterV} and \eq{eq:fit_clusterM} describe the distribution
of the topological defects as the sum of
an ultraviolet component (first, inverse power term) and
an infrared component (second, Gaussian term).

\begin{figure*}[!htb]
\begin{center}
\begin{tabular}{cc}
\includegraphics[scale=0.34,clip=true]{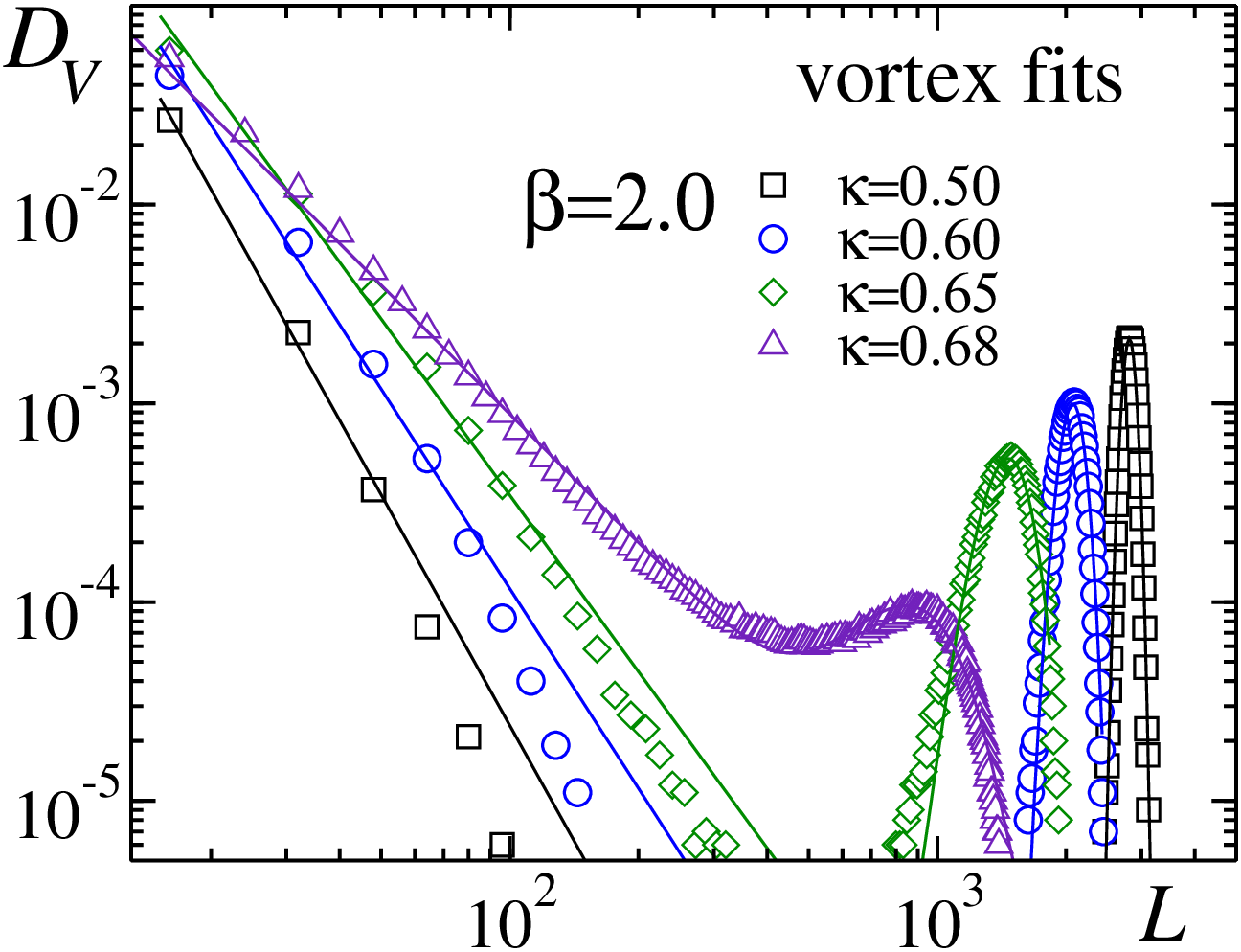} \hspace{5mm} &
\includegraphics[scale=0.34,clip=true]{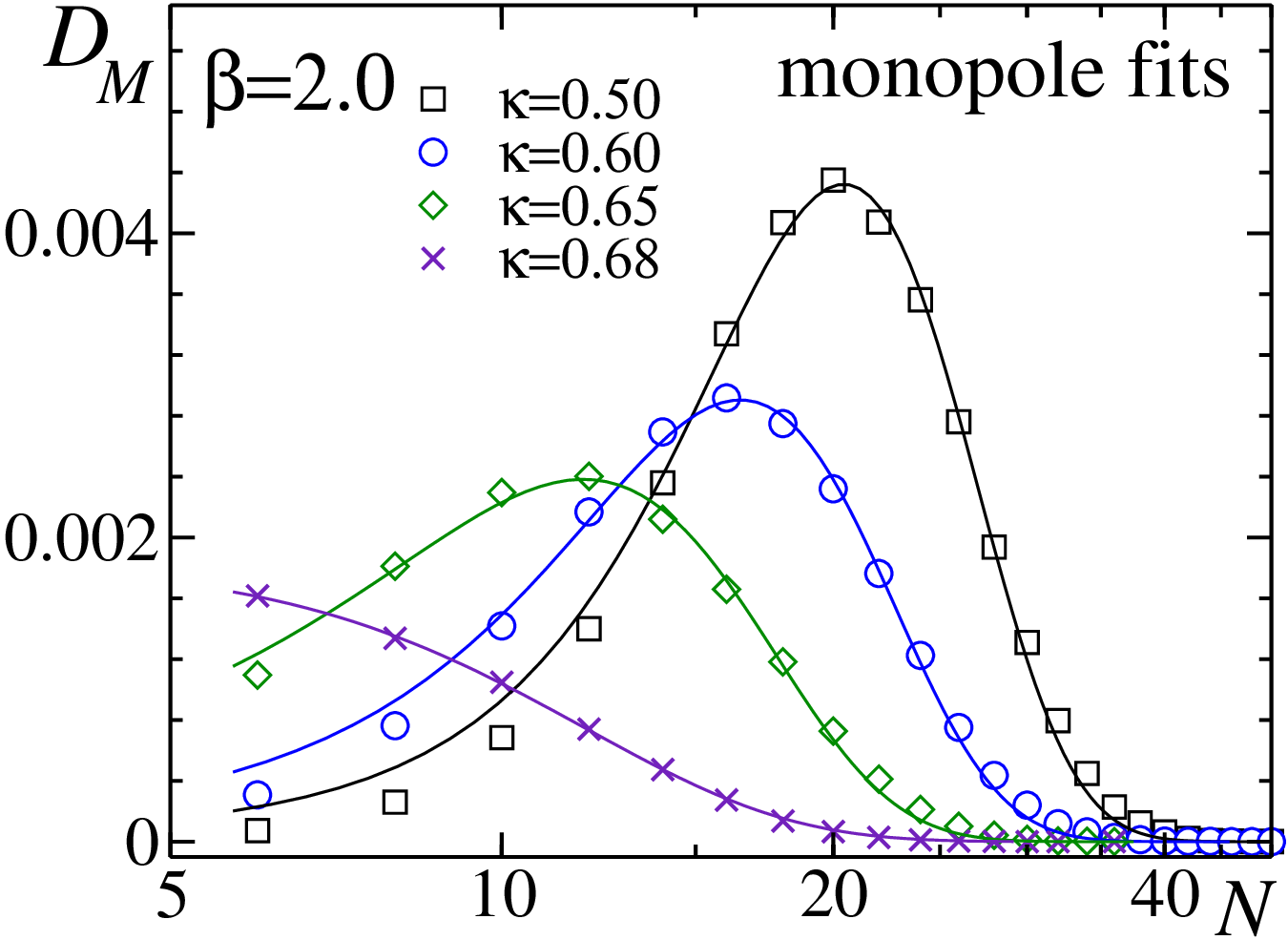} \\
  (a) & \hspace{5mm} (b)
\end{tabular}
\end{center}
\vskip -5mm
\caption{Fits of (a) the monopole and (b) the vortex histograms by
Eqs.~\eq{eq:fit_clusterV} and \eq{eq:fit_clusterM}.}
\label{fig:bclusters_beta_fit}
\end{figure*}

As we have seen, the second (infrared) peaks in the monopole number
and vortex length distributions
are absent in the deconfinement phase. Therefore we fit the distributions by
Eqs.~\eq{eq:fit_clusterV} and \eq{eq:fit_clusterM} only in the confinement phase.
The fits of the vortex length
distributions are shown in Figure~\ref{fig:bclusters_beta_fit}(a).
One can
see that the fitting function describes well
the data both in the low-$L$ and high-$L$ regions.
To fit the monopole distributions we have set $C_{1,M}=0$ since there is not enough
data to fix the ultraviolet part of the distribution. We have performed the fit
for $N\geqslant 6$ as it is shown in Figure~\ref{fig:bclusters_beta_fit}(b).

\begin{figure*}[!htb]
\begin{center}
\includegraphics[scale=0.34,clip=true]{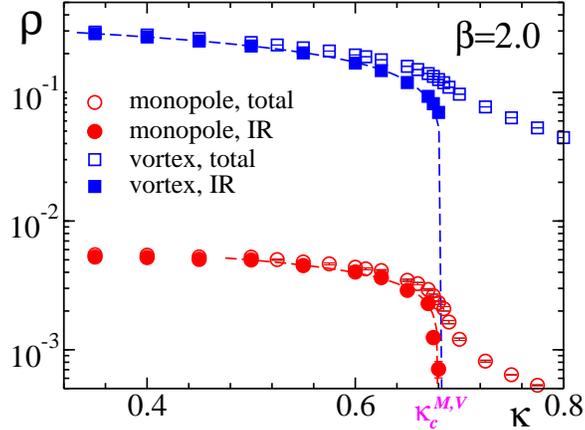}
\end{center}
\vskip -5mm
\caption{The densities of topological defects in infrared clusters as given by
fits~\eq{eq:fit_clusterV} and \eq{eq:fit_clusterM}.
The fits by Eq.~\eq{eq:IRdens_fit} are shown by dashed lines.}
\label{fig:bdensity_IR}
\end{figure*}

The densities of vortices and monopoles in infrared clusters are shown in
Figure~\ref{fig:bdensity_IR} as a function of $\kappa$ in the confining phase.
One observes that the density of the infrared part of both
types of topological defects drops down
when the phase transition point is approached,
in agreement with our expectations.
Deeper in the
confining phase the infrared densities are practically
saturating the total densities.

We have fitted the infrared densities of both types of topological defects
by functions of the form
\beqn
\rho^{V,M}_{IR} (\kappa)= B_{V,M} \cdot (\kappa_c^{V,M} - \kappa)^{\gamma^{V,M}}\,, \qquad \kappa < \kappa_c\, ,
\label{eq:IRdens_fit}
\eeqn
assuming a ``critical behavior'' near the phase transition.
Both fits are shown in Figure~\eq{fig:bdensity_IR} as dashed lines. The
transition points are:
\beqn
\kappa^{M}_c = 0.680(3)\,, \qquad \kappa^{V}_c = 0.684(1)
\eeqn
from the infrared
monopole and vortex densities,
respectively. They are in agreement with each other as well as with the values
determined by from the susceptibilities of the Polyakov loops and the vortex density.
The ``critical exponents'' are different:
$\gamma^{M} = 0.28(2)$ and $\gamma^{V} = 0.37(1)$.

\begin{figure*}[!htb]
\begin{center}
\includegraphics[scale=0.34,clip=true]{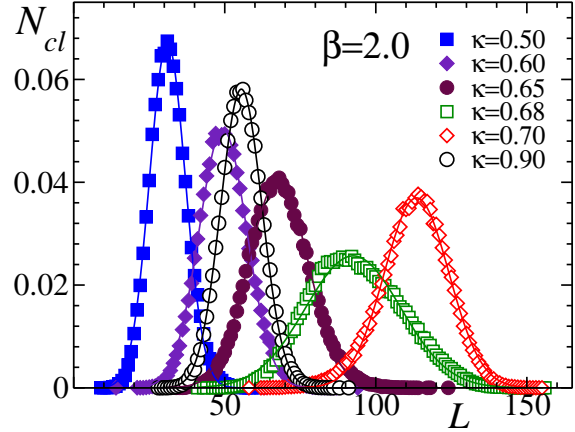}
\end{center}
\vskip -5mm
\caption{The multiplicity distributions of vortex clusters and the corresponding
fits by the Gaussian distribution~\eq{eq:multi_fit} for various $\kappa$ values.}
\label{fig:clust_mult}
\end{figure*}

Yet another interesting aspect of the cluster structure is the cluster
multiplicity distribution $D(N_{cl})$ which represents the total number of vortex
clusters per configuration (irrespective of the cluster size). We show
$D(N_{cl})$ for various values of hopping parameter
$\kappa$ in Figure~\ref{fig:clust_mult}.
The distribution of the cluster number can be
well described by a Gaussian function,
\beqn
D(N_{cl}) = A \exp\Bigl\{ - \frac{(N_{cl} - \bar{N}_{cl})^2}{2 \Lambda} \Bigr\}\,.
\label{eq:multi_fit}
\eeqn
The fits are shown in Figure~\ref{fig:clust_mult} as solid lines.

\begin{figure*}[!htb]
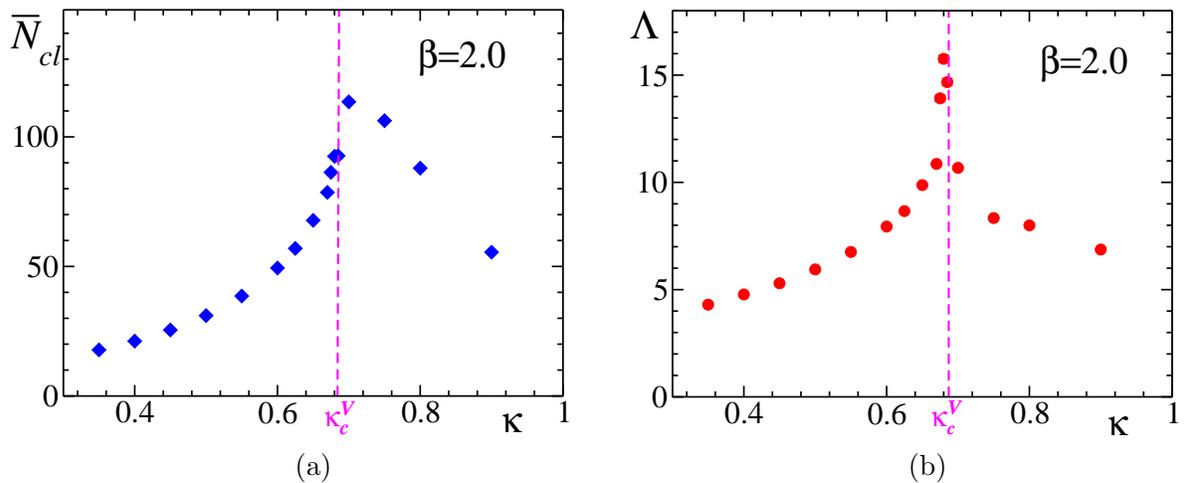

\begin{center}
\begin{tabular}{cc}
\includegraphics[scale=0.34,clip=true]{bclustcenter.eps} \hspace{5mm} &
\includegraphics[scale=0.34,clip=true]{bclustwidth.eps} \\
  (a) & \hspace{5mm} (b)  \\
\end{tabular}
\end{center}
\vskip -5mm
\caption{(a) The average number $\bar{N}_{cl}$ of clusters and (b)
the width $\Lambda$ of the multiplicity distribution as functions of $\kappa$.}
\label{fig:clust_mult_parameters}
\end{figure*}

The fitted average cluster multiplicity $\bar{N}_{cl}$ and
the width of the distribution $\Lambda$ are presented in
Figure~\ref{fig:clust_mult_parameters} as functions of $\kappa$.
Both parameters are peaked near the phase transition indicating that at the critical
value of the hopping parameter the few large infrared clusters -- existing in
the confinement phase -- decompose into many smaller clusters as the system goes over
into the deconfinement phase. According to Figure~\ref{fig:bdefect_susc}(a) the
monopole and vortex densities drop slightly with increasing $\kappa$ at $\kappa_c$.
Therefore, the increase of the cluster number is associated predominantly with a
cluster--restructuring process and not with a cluster creation phenomenon.
Moreover, the increase of the width $\Lambda$ indicates the existence
of large fluctuations in the vortex system at $\kappa \approx \kappa_c$.

\subsection{The zero-temperature case at fixed hopping parameter $\kappa=1.0$}
\label{subsec:T0_at_kappa1.0}

\begin{figure*}[!htb]
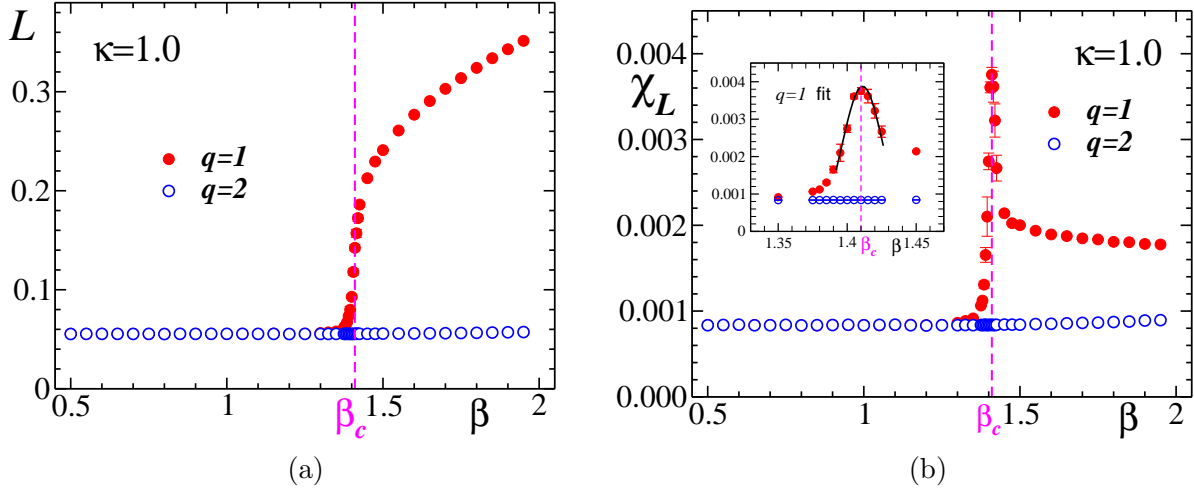

\begin{center}
\begin{tabular}{cc}
\includegraphics[scale=0.34,clip=true]{kpolyakov.loop.eps} \hspace{5mm} &
\includegraphics[scale=0.34,clip=true]{kpolyakov.susc.eps} \\
  (a) & \hspace{5mm} (b)  \\
\end{tabular}
\end{center}
\vskip -5mm
\caption{(a) The expectation values of the Polyakov loop and (b) their
susceptibilities {\it vs.} $\kappa$. The insert in Figure (b) shows the
behavior of the $q=1$ susceptibility in the vicinity of the phase
transition together with its fit.}
\label{fig:kpolyakov}
\end{figure*}

In this subsection we map out the phase structure along an horizontal line
through the phase diagram of Fig.~\ref{fig:phase_diagram_general}.
At $\kappa$ as large as $\kappa=1.0$ the confinement phase is realized only
for relatively small $\beta$.
In Fig.~\ref{fig:kpolyakov}
we show the {\it v.e.v.}'s of the $q=1$ and $q=2$
Polyakov loops {\it vs.} $\beta$. Only the susceptibility of the
$q=1$ Polyakov loop shows a peak which leads to an estimate
\beqn
  \beta_c \equiv \beta_c^{q=1}=1.410(2)
  \label{eq:kappa_c:polyakov}
\eeqn
using a fitting function $\chi^{\rm fit}(\beta)$ analogously to (\ref{eq:fitting_function}).
The position of the phase transition
defined from  the fit of the $q=1$ Polyakov loop
is shown by a vertical dashed line.

\begin{figure*}[!htb]
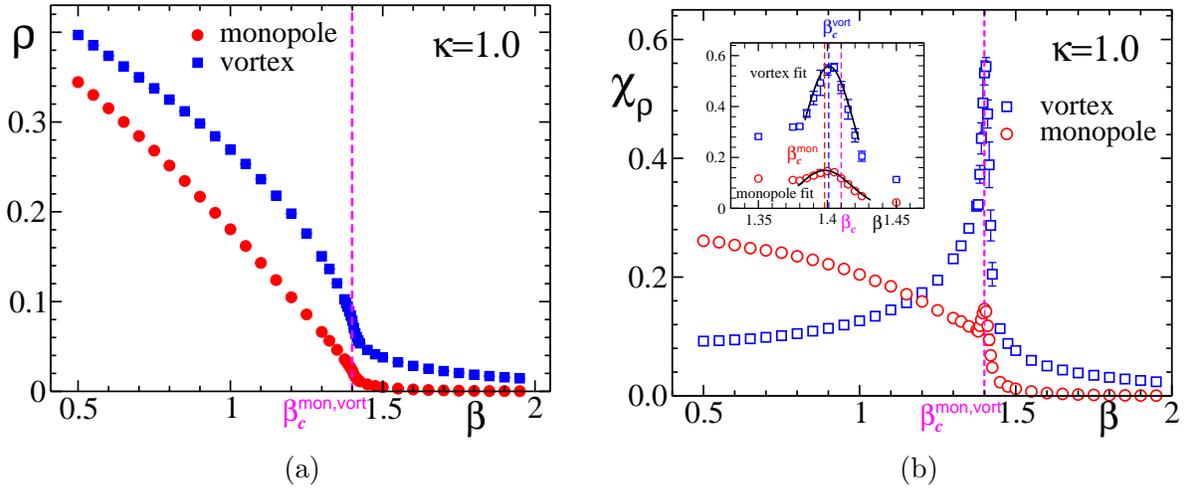

\begin{center}
\begin{tabular}{cc}
\includegraphics[scale=0.34,clip=true]{kdefect.dens.eps} \hspace{5mm} &
\includegraphics[scale=0.34,clip=true]{kdefect.susc.eps} \\
  (a) & \hspace{5mm} (b)  \\
\end{tabular}
\end{center}
\vskip -5mm
\caption{(a) The densities of the topological defects and (b) their
susceptibilities {\it vs.} $\beta$.
The insert in Figure (b) shows the behavior of the monopole and vortex susceptibilities in
the vicinity of the phase transition
with their fits and critical $\beta$'s.}
\label{fig:kdefect}
\end{figure*}

In the confining phase we have now a density of monopoles comparable in
magnitude to the density of vortices. In Fig.~\ref{fig:kdefect}
we show both densities
and the respective susceptibilities. Values as $\beta=2.0$ which have led to a
vortex--driven transition described
in the previous subsection are now belonging to the deconfinement phase.
The peak of the susceptibility of the vortex density is still
more pronounced at the transition, but now the susceptibility of the monopole
density also shows a little peak.

The positions of the peaks for the monopole and vortex susceptibilities are slightly
shifted compared to the peak of the $q=1$ Polyakov loop susceptibility,
Fig.~\ref{fig:kpolyakov}. The fits of the monopole and vortex susceptibilities
in the vicinity of the phase transition give the
following values for $\beta_c$ corresponding to the transition, respectively:
\beqn
  \beta_c^{\mathrm{mon}} = 1.398(1)\,,\qquad \beta_c^{\mathrm{vort}} = 1.401(1)\,.
  \label{eq:kappa_c:defects}
\eeqn
One can see that the couplings $\beta_c$ obtained with the help of the monopole
and vortex susceptibilities are very similar to each other.

\begin{figure*}[!htb]
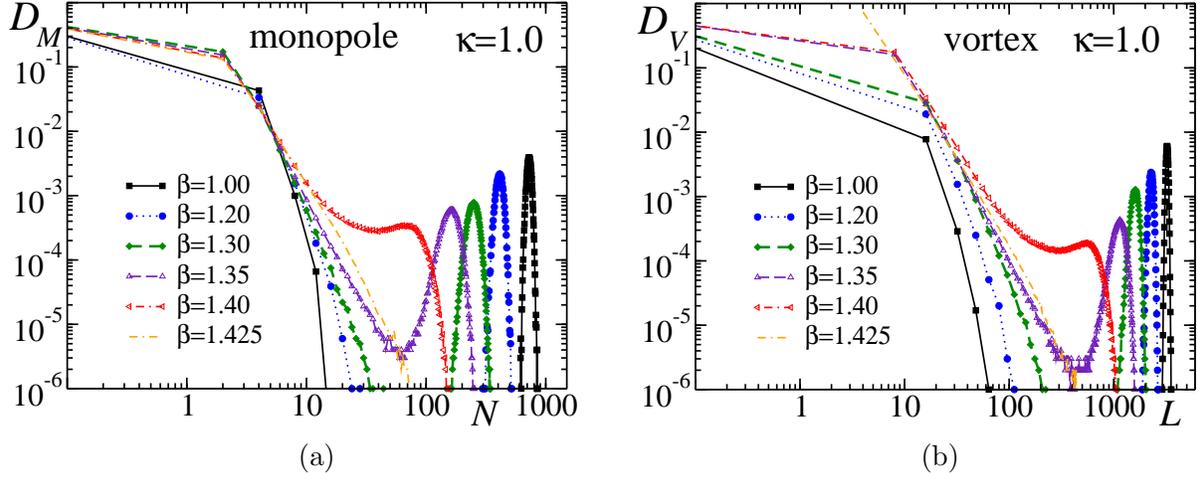

\begin{center}
\begin{tabular}{cc}
\includegraphics[scale=0.34,clip=true]{kmonhist.eps} \hspace{5mm} &
\includegraphics[scale=0.34,clip=true]{kvorthist.eps} \\
  (a) & \hspace{5mm} (b)  \\
\end{tabular}
\end{center}
\vskip -5mm
\caption{Histograms of the (a) monopole and (b) vortex distributions.}
\label{fig:khist}
\end{figure*}

\begin{figure*}[!htb]
\begin{center}
\begin{tabular}{cc}
\includegraphics[scale=0.34,clip=true]{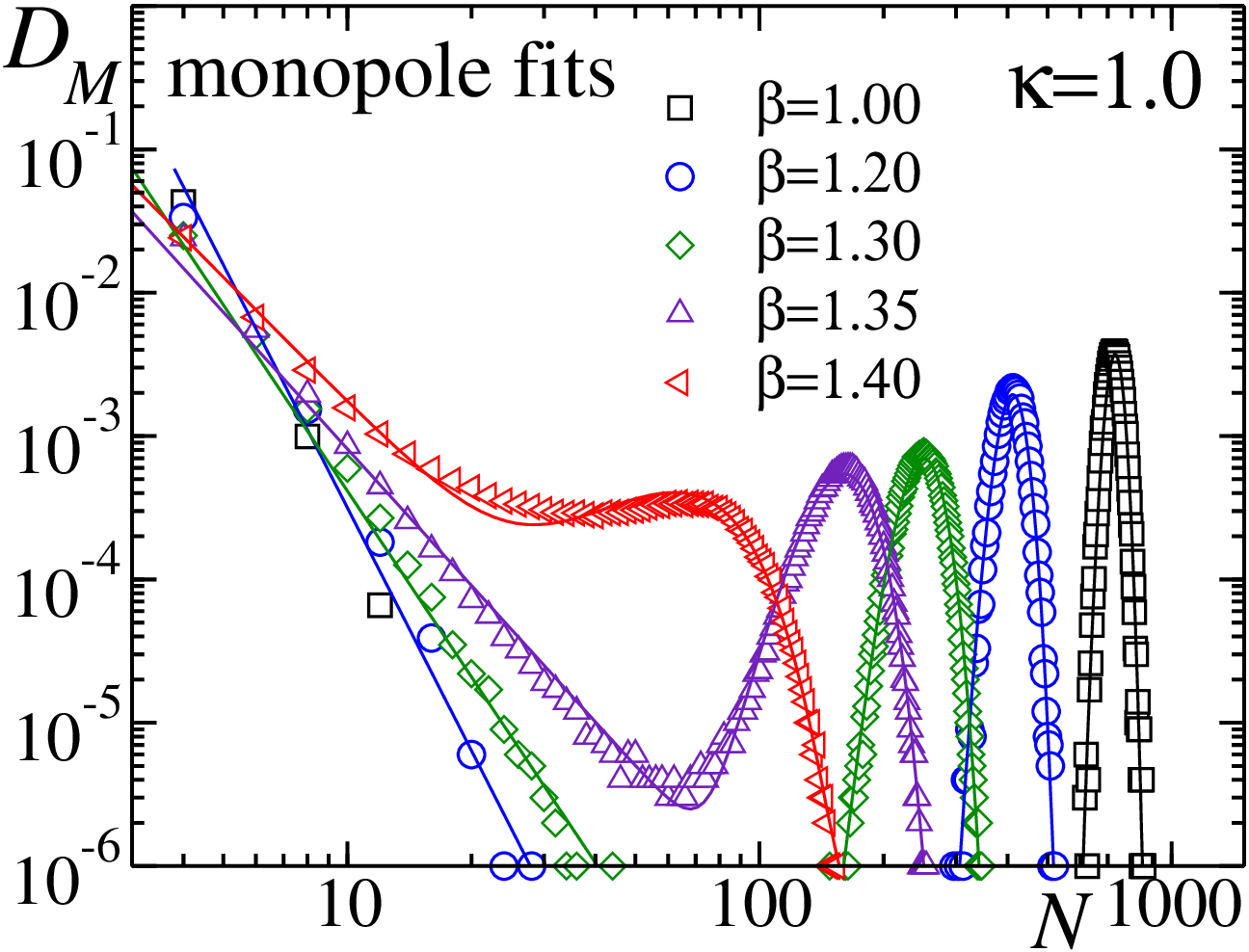} \hspace{5mm} &
\includegraphics[scale=0.34,clip=true]{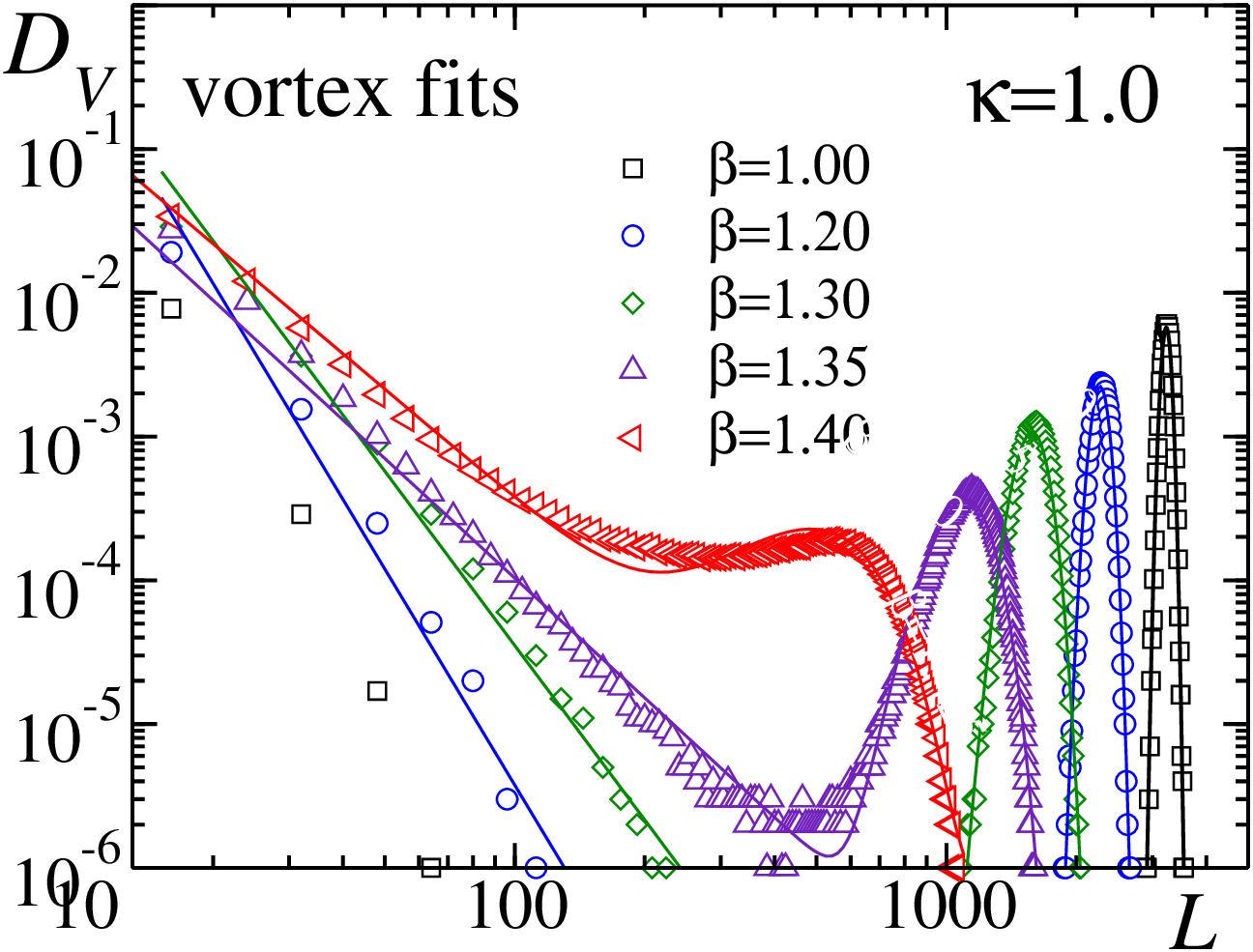} \\
  (a) & \hspace{5mm} (b)
\end{tabular}
\end{center}
\vskip -5mm
\caption{Fits of (a) the monopole and (b) the vortex histograms
by Eqs.~\eq{eq:fit_clusterV} and \eq{eq:fit_clusterM}.}
\label{fig:clusters:kappa:fit}
\end{figure*}

Histograms of the vortex length an monopole number distribution for the
individual vortex clusters are shown in Fig.~\ref{fig:khist}
for various $\beta$
values near $\beta_c$. Fits corresponding to the two-component forms
(\ref{eq:fit_clusterV}) and (\ref{eq:fit_clusterM}) are presented
in Fig. \ref{fig:clusters:kappa:fit}.

\begin{figure*}[!htb]
\begin{center}
\includegraphics[scale=0.34,clip=true]{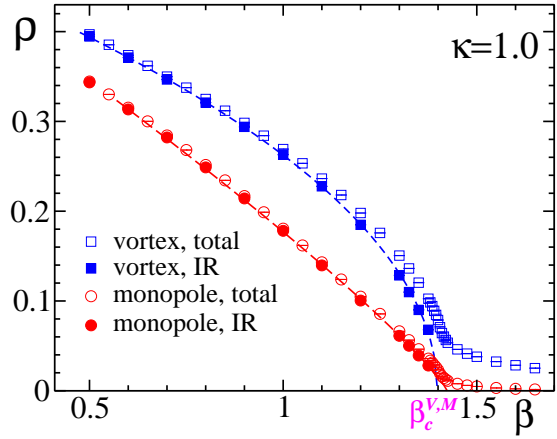}
\end{center}
\vskip -5mm
\caption{The densities of the topological defects in infrared clusters
as given by fits~\eq{eq:fit_clusterV} and \eq{eq:fit_clusterM}.
The fit curves using $\rho^{M,V}_{IR}(\beta)$
similar to Eq.~\eq{eq:IRdens_fit} are shown by dashed lines.
}
\label{fig:kdefect:IR}
\end{figure*}

The fitted values of the infrared monopole and vortex densities are
shown in Fig. \ref{fig:kdefect:IR}
together with the total densities
as functions of $\beta$.
Again deeper in the
confining phase the infrared densities practically
saturate the total densities.
The respective $\beta_c$--values marking the transition
from the infrared clusters are found fitting the $\beta$-dependence with an ansatz
$\rho^{M,V}_{IR}(\beta)$
similar to Eq.~\eq{eq:IRdens_fit}.
The fit curves are shown in the Figure as well.
The corresponding critical values extracted from the
infrared monopole and the vortex densities are:
\beqn
  \beta^{M}_c & = & 1.424(3)\,, \qquad \beta^{V}_c = 1.399(2)\,.
  \label{eq:crit:beta:IR}
\eeqn
The ``critical exponents'' are rather different again:
$\gamma^{M}= 0.86(1)$ and $\gamma^{V} = 0.499(4)$.
The deviation between the values of the
transition points $\beta_c$ obtained from fitting the two
infrared densities~\eq{eq:crit:beta:IR} and from the $\beta_c$ values
obtained from the
defect density susceptibilities~(\ref{eq:kappa_c:defects})
is a finite volume effect, caused presumably by an enhanced sensitivity of
these densities to the volume of the system.

\section{The Landau gauge photon propagator at the phase transition}
\label{sec:phase_structure_photon}

Here we want to clarify how the behavior of the photon propagator
reflects the properties of the two phases. We restrict ourselves to the case of
zero temperature (symmetric lattice).
In order to calculate the (gauge dependent) photon propagator directly,
a special gauge has to be chosen. In our case this is the minimal Landau gauge.
This gauge is defined by finding the global maximum of the gauge functional
\beqn
F=\frac{1}{N_{\rm links}}\sum_{x,\mu} \cos(\theta^G_{x,\mu})
                                      \rightarrow {\rm max}
\label{def:Landau_gauge}
\eeqn
with respect to gauge transformations $G$.
Details of the gauge fixing procedure can be found in
Ref.~\cite{Chernodub:2004yv}.

The photon propagator in lattice momentum space is the ensemble average
over gauge-fixed configurations
of the following bilinear in the Fourier transformed lattice gauge potential $\tilde{A}$
(with $p_\mu(k_\mu)=  (2/a) \sin (\pi k_\mu/L_\mu)$)
\beqn
D_{\mu\nu}({\vec p}) = \langle \tilde{A}_{ {\vec k},\mu}
                               \tilde{A}_{-{\vec k},\nu} \rangle \,.
\label{def:propagator}
\eeqn
constructed out of the gauge fixed links:
\beqn
\tilde{A}_{{\vec k},\mu} =
\left(\frac{1}{L^3}\right)^{1/2}
\sum\limits_x
\exp \Bigl( 2 \pi i~\sum_{\nu=1}^{3} \frac{k_{\nu}
(~x_{\nu}+\frac{1}{2}\delta_{\nu\mu}~) } {L_{\nu}} \Bigr)
~ \sin  \theta^G_{x,\mu} \,, \quad
k_\mu=0, \pm 1,..., \pm \frac{L_\mu}{2} \,.
\label{def:fourier_transformation}
\eeqn
The lattice equivalent of the same continuum $p^2$ can be realized
by different vectors $\vec k$ which eventually
would reveal a breaking of rotational invariance on the lattice.

We consider the general tensor structure with two form factors $D$ and $F$ for the propagator:
\beqn
   D_{\mu\nu}({\vec p})= \left(\delta_{\mu\nu}- \frac{p_\mu p_\nu}{p^2}\right) D(p^2)
   + \frac{p_\mu p_\nu}{p^2} \frac{F(p^2)}{p^2} \,.
\label{def:tensor_structure}
\eeqn
When the Landau gauge is exactly fulfilled, $F(p^2) \equiv 0$.
On the lattice, this is actually the case as soon as one of the local maxima of
the gauge functional (\ref{def:Landau_gauge}) is reached.
Since $D_{\mu\nu}({\vec p})$ is only approximately rotationally
invariant, the form factor $D(p^2)$   scatters instead of forming a smooth
function of $p^2$.

We have  measured the propagator on a lattice of size $40^3$
at $\beta=2.0$ as function of the hopping parameter $\kappa$.
O(50) independent configurations have been considered outside the phase transition region and
O(200) (roughly) independent configurations near the second order transition.
The autocorrelation times have been estimated by looking both at the monopole and
vortex densities, $\rho_{\rm mon}$ and $\rho_{\rm vort}$. In order to fight with huge
autocorrelations, we separated the used configurations by 10000 Monte Carlo sweeps.

As in the case of  the $Q=1$  AHM  we parameterize the function $D(p^2)$ by
\beqn
  D(p^2) = \frac{Z\,m^{2 \alpha}}
  {\beta\left[p^{2 (1+\alpha)} + m^{2 (1+ \alpha)} \right]} + C\,,
  \label{eq:fit:total}
\eeqn
where $Z$, $\alpha$, $m$ and $C$ are fitting parameters.
Their meaning  is as follows: $Z$ is
the renormalization of the photon wavefunction, $\alpha$ is the
anomalous dimension, $m$ is a mass parameter.
The parameter $C$ corresponds to a $\delta$-like interaction in the coordinate
space and, consequently, is irrelevant for long-range physics.
{}From previous studies we know that this fit ansatz well describes the data in pure
confining and Higgs phases separated by a crossover or a first order phase transition.

\begin{figure}[!htb]
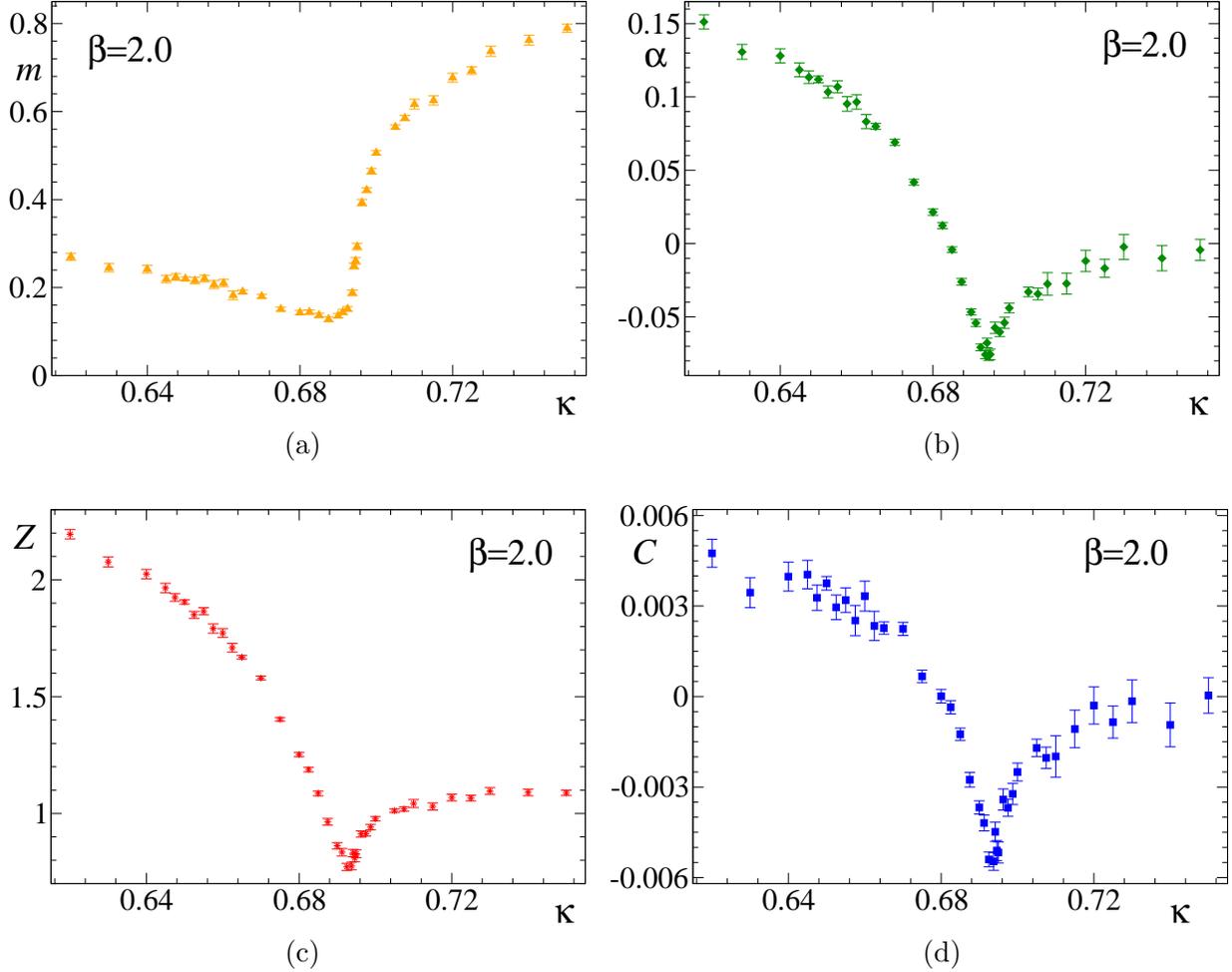

  \begin{center}
  \begin{tabular}{cc}
\includegraphics[scale=0.33,clip=true]{m.eps} \hspace{5mm}&
\includegraphics[scale=0.33,clip=true]{alpha.eps} \\
    (a) & \hspace{5mm} (b)\\ \\
\includegraphics[scale=0.33,clip=true]{Z.eps} \hspace{5mm}&
\includegraphics[scale=0.33,clip=true]{C.eps} \\
  (c) & \hspace{5mm} (d)
  \end{tabular}
  \end{center}
  \caption{The fitting parameters for the propagator measured on $40^3$ at
  $\beta=2.0$ vs. $\kappa$.
  }
  \label{fig:fits}
\end{figure}

Treating now the same ansatz, we present in Figs.~\ref{fig:fits} the obtained fit parameters
as function  of the hopping parameter $\kappa$.
The optimal fit parameters have been found by minimizing $\chi^2/d.o.f$.
Outside the critical region ($\kappa_c \approx 0.68$ -- $0.70$) the propagator data are perfectly
described by the simple fit ansatz~(\ref{eq:fit:total}).
The quality of the fits is the same as in the $Q=1$ AHM~\cite{Chernodub:2004yv},
therefore we do not show a comparison of the measured data with the fits ansatz.

The propagator fit leads to a minimum of the photon mass (Fig.~\ref{fig:fits}(a)). On the Higgs side of the
transition the photon acquires its mass via the Higgs mechanism.
The noted minimum could be explained similar to the $Q=1$ model via
the interference between perturbative mass and
Debye mass effects.

As shown in Fig.~\ref{fig:fits}(b) the anomalous dimension $\alpha$ drops towards $\kappa_c$
and becomes negative near criticality within the used ansatz.
It tends to zero from below at larger $\kappa$, deeper in the Higgs region.
The strong wave function renormalization effects in the confining phase ($Z$ is much larger
than one, see Fig.~\ref{fig:fits}(c)) weaken
towards the transition and $Z$ approaches a roughly constant value
near one, deeper in the Higgs phase.
For completeness, the contact term $C$ as function of $\kappa$ is shown in
Fig.~\ref{fig:fits}(d) as well.

In the critical region of the second order phase transition
we observed that the propagator data for the smallest available
lattice momenta seem to indicate a non-vanishing $p^2 D(p^2)$ for $p^2 \to 0$,
contrary to the used fit ansatz (as long as the {\it{fitted}} mass remains finite).
For the worst observed case at $\kappa=0.695$ the smallest measured value  for $p^2 D(p^2)$ lies
roughly two times above the fit curve (at $p^2\approx 0.025$).
The noticed deviation between data and fit curve is exclusively related only to the smallest
lattice momenta.
Nevertheless, the obtained values for $\chi^2/d.o.f.$ do not exceed 1.17 and do not differ
significantly from the $\chi^2$ values outside the transition.
More sophisticated parameterizations for the propagator and eventually measurements
for still smaller momenta (larger lattices) are needed to clarify this problem
what also might influence the actual values of our fit parameters.
This, however, was beyond the scope of our present investigation.

\section{Conclusions}
\label{sec:Conclusions}

Summarizing, we have found that the compact Abelian Higgs model with doubly-charged
matter field has a rich structure in terms of topological defects: monopoles and vortices.
This structure has been outlined first by analytic considerations in the London
limit and in the Villain representation. In particular, the topological
character of confinement for $mod(q,Q) \ne 0$ via the Aharonov-Bohm
(linking number) interaction has been pointed out.

Extending first results, obtained for a suitable finite-temperature regime
of the model and published already in \cite{ref:U1Q2PLB}, we could show by
numerical simulation in the London limit, that the main property of the model,
the existence of vortex clusters with monopoles attached to them is a generic
feature of the model. This structure can explain that in the confining corner
of the $\beta$-$\kappa$ phase diagram only odd-$q$ external charges are confined while
even-$q$ charges suffer string breaking.

In the confinement phase we found that
the vortex clusters are percolating. Both the length distribution of the vortex
clusters and the monopole number distribution kept inside the clusters
give rise to the definition of two order parameters, the
``infrared densities'' of monopoles and vortices.
The infrared (percolating) component of the vortex network
and the corresponding order parameters vanish in the Higgs phase.

For not too large lattice size at $T=0$
the phase transition is signalled also
by the original gauge field variables, using for example the
susceptibilities of the $q=1$ Polyakov loop on finite lattices (directly
related to the loss of confinement for charge $q=1$ external charges).
More quantitative description one can get from
the susceptibilities of the monopole and vortex densities, and finally
in terms of the two infrared order parameters. Different
parts of the confining phase (high $\beta$ and small $\kappa$ on one hand and
small $\beta$ and high $\kappa$ on the other) may differ in the detailed mechanism,
being vortex driven (at low $\kappa$) or monopole driven (at low $\beta$).

We also have shown that the gauge boson propagator changes its character at
the phase transition to the Higgs phase where it changes from confining to
Yukawa type. There are remarkable differences from the $Q=1$ AHM.
Using our simple parametrization (\ref{eq:fit:total}) for the form factor
$D(p^2)$ we found that in the $Q=2$ case both the critical dimension $\alpha$, the photon
renormalization function $Z$ and the contact term $C$ go through minima at
the phase transition point, which is not the case for the $Q=1$ case.
Moreover, in a vicinity of the phase transition the renormalization
parameter $Z$ gets smaller than unity and the propagator
can be described by a negative anomalous dimension
contrary to the $Q=1$ case.
We were not able to see the fitted gauge boson mass vanishing at pseudocriticality
(eventually to be expected from the second order nature of the phase transition).
So, this interesting behavior can be confirmed only using larger lattice sizes
to access still lower lattice momenta and more sophisticated parameterizations
for the propagator.

This study deepens our understanding of the phase transition in this model.
It also further develops the paradigm for the cooperative role that monopoles
and vortices play in non-Abelian gluodynamics.
The physical picture -- observed in the present study -- has a close analogy with
QCD (in 4D) where tight correlations between Abelian monopoles and center vortices
have already been observed. The non-diagonal gluons are usually ignored
after the Abelian projection (from the maximally Abelian gauge). In fact, they
are charge-two matter field with respect to the remaining Abelian gauge symmetry.
If these matter fields would be retained as dynamical, this might naturally lead
to the formation of monopole sheets on vortex sheets, which in turn should be
responsible for the permanent (in the pure gauge model) confinement of the fundamental
charges (quarks) and for the flattening of the potential between the adjoint charges
(``static gluons'').

\begin{acknowledgments}
M.N.Ch. is supported by grants RFBR 04-02-16079 and MK-4019.2004.2. E.-M. I. is supported
by DFG (FOR 465/Mu 932/2).

\end{acknowledgments}

\end{document}